\documentclass{article}

   \usepackage[nonatbib]{neurips_2023_ml4ps}

\usepackage[utf8]{inputenc} 
\usepackage[T1]{fontenc}    
\usepackage{hyperref}       
\usepackage{url}            
\usepackage{booktabs}       
\usepackage{amsfonts}       
\usepackage{amsmath}
\usepackage{nicefrac}       
\usepackage{microtype}      
\usepackage{xcolor}         
\usepackage{graphicx}
\usepackage{subcaption}
\usepackage{tabularx}
\usepackage{multirow}
\usepackage{tikz}
\usetikzlibrary{tikzmark}
\usetikzlibrary{positioning, shapes.geometric, arrows.meta, fit}

\title{Redefining Super-Resolution: Fine-mesh PDE predictions without classical simulations}

\author{%
  Rajat Kumar Sarkar\thanks{Corresponding author}\\
  Researcher \\
  TCS Research \\
  \texttt{rajat.sarkar1@tcs.com}\\
  \And
  Ritam Majumdar \\
  Researcher \\
  TCS Research \\
  \texttt{ritam.majumdar@tcs.com} \\
  \AND
  Vishal Jadhav \\
  Scientist \\
  TCS Research \\
  \texttt{vi.suja@tcs.com} \\
  \And
  Sagar Srinivas Sakhinana \\
  Scientist \\
  TCS Research \\
  \texttt{sagar.sakhinana@tcs.com} \\
  \And
  Venkataramana Runkana \\
  Chief Scientist \\
  TCS Research \\
  \texttt{venkat.runkana@tcs.com} \\
}

\begin{document}

\maketitle

\begin{abstract}
  In Computational Fluid Dynamics (CFD), coarse mesh simulations offer computational efficiency but often lack precision. Applying conventional super-resolution to these simulations poses a significant challenge due to the fundamental contrast between downsampling high-resolution images and authentically emulating low-resolution physics. The former method conserves more of the underlying physics, surpassing the usual constraints of real-world scenarios. We propose a novel definition of super-resolution tailored for PDE-based problems. Instead of simply downsampling from a high-resolution dataset, we use coarse-grid simulated data as our input and predict fine-grid simulated outcomes. Employing a physics-infused UNet upscaling method, we demonstrate its efficacy across various 2D-CFD problems such as discontinuity detection in Burger's equation, Methane combustion, and fouling in Industrial heat exchangers. Our method enables the generation of fine-mesh solutions bypassing traditional simulation, ensuring considerable computational saving and fidelity to the original ground truth outcomes. Through diverse boundary conditions during training, we further establish the robustness of our method, paving the way for its broad applications in engineering and scientific CFD solvers.
\end{abstract}

\section{Introduction}
Computational Fluid Dynamics (CFD) plays a crucial role in comprehending intricate physical phenomena spanning scientific and engineering domains, including aerospace \cite{raj1995requirements}\cite{singh2010computational}, automotive \cite{szewc2018gpu}, energy\cite{miller2013review}, and more. To gain a profound understanding of these physical phenomena, it becomes imperative to conduct simulations at high mesh resolutions of the governing equation of fluid flow like the Navier-Stokes equation, encompassing a broad spectrum of fluid structures, ranging from large-scale patterns\cite{merzari2017large}\cite{liu2008turbulence} to subgrid-scale features like small eddies within fluid flow systems \cite{mason1986magnitude}\cite{sarlak2015role}. However, simulation of fluid flow within intricate geometries at high mesh resolutions is inherently computationally intensive and time-consuming\cite{roelofs2019cfd}. This has led to the popularity of coarse mesh simulations, primarily due to their computational efficiency. Nevertheless, coarse mesh simulations deal with a persistent challenge - their inherent limitation of low mesh resolution, often compromising the precision of the results obtained. Addressing this resolution disparity becomes paramount in accurately capturing the subtleties of fluid dynamics. 

In order to circumvent the issue of simulating high-resolution mesh solutions of PDEs, several researchers have adapted the super-resolution technique, commonly used in Computer Vision\cite{irani1991improving}\cite{bannore2009iterative}\cite{tian2011survey} to map low-resolution data to high-resolution data using a supervised loss. This approach involves an initial downsampling of the original high-resolution data to low resolution, accomplished through various downsampling techniques such as max-pooling, average-pooling, and nearest-neighbor methods. Subsequently, they leverage a combination of numerous variants of deep learning architectures, including multi-layer perceptron \cite{erichson2020shallow, nair2020leveraging}, convolutional neural networks \cite{fukami2019super, obiols2021surfnet, liu2020deep, zhou2022neural, esmaeilzadeh2020meshfreeflownet}, and generative adversarial networks \cite{xie2018tempogan, bode2019deep, kim2021unsupervised, guemes2021coarse, yousif2021high, bode2021using}. These techniques are employed to reconstruct the original high-resolution data from the downsampled low-resolution data they have generated. Moreover, some researchers have integrated physics-based principles into their networks to effectively capture the underlying physics within the low-resolution data, enhancing the reconstruction of high-resolution data \cite{gao2021super, bode2019deep, bode2021using, esmaeilzadeh2020meshfreeflownet, arora2022physrnet}.

However, high-resolution mesh solutions of partial differential equations (PDEs) contain an extensive amount of information regarding the underlying physics within the mesh stencil. In contrast, low-resolution mesh stencils tend to exhibit irregularities in capturing these governing physics. Applying the conventional definition of super-resolution directly to the world of PDEs presents two notable challenges. Firstly, the process of generating a dataset by downsampling high-resolution data to create low-resolution data differs significantly from generating simulations of low-resolution data using conventional numerical solvers. Secondly, there is a shortage of high-resolution mesh data, making it difficult to acquire sufficient training data for super-resolution models. This distinction arises from the complex physical properties that are inherently associated with high-resolution data. Simply downsampling from high-resolution data results in retaining the majority of the governing physics(refer to \ref{sec:Downsampling}), which may not accurately reflect real-world scenarios. To address this challenge, we have devised an alternative approach for framing the super-resolution problem to align more closely with practical applications. Instead of downsampling high-resolution data, we utilize coarse-grid simulated data as input and predict fine-grid simulated data as the super-resolution output. This approach emulates real-world conditions accurately and overcomes the limitations of conventional super-resolution techniques. 

In summary, our physics-informed UNet upscaling approach effectively predicts fine mesh data features from coarse mesh data, demonstrating its effectiveness across various CFD datasets. Enabling the generation of high-fidelity fine-mesh solutions without the need for traditional time-consuming CFD simulations, our method offers both significant time savings and accuracy. We've demonstrated its robustness through diverse boundary conditions during training, highlighting its potential for widespread applications in engineering and scientific CFD solvers. 

\section{Methodology}
\subsection{Problem formulation}
Consider a training dataset consisting of pairs of coarse mesh data and fine mesh data represented as {($C, F$)}. The primary goal of this research is to train an inductive supervised model ($f: C\rightarrow F$) to effectively capture and map the non-linear relationship between coarse mesh data $C(x_1,x_2,...x_n) \in R^{m}$ and fine mesh data $F(x_1,x_2,...x_n) \in R^{d}$ of $n$ features, where $m \ll d$. Mapping learns the characteristics of the system variables and properties described by governing equations. The model takes coarse mesh data at a given time step $t$  and predicts the corresponding fine mesh data, effectively performing upsampling across features ($x_n$) as shown in table \ref{tab:Cost}.
\subsection{Model Architecture}
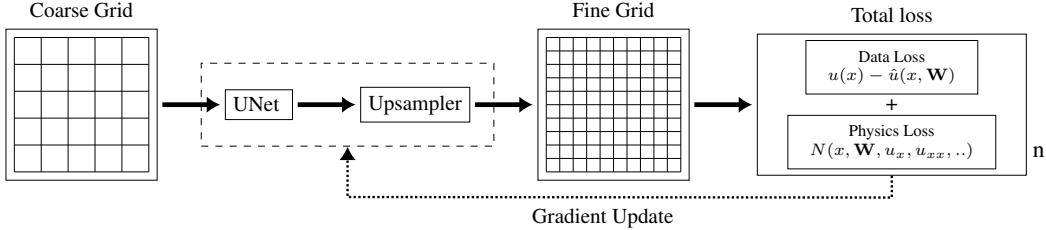
\begin{figure}[htbp]
    \centering
    \resizebox{\textwidth}{!}{
        \begin{tikzpicture}[node distance=0.3cm,font=\small,baseline=(current bounding box.center)]
                \def\numRows{5}
                \def\numCols{5}
                
                \def\cellSize{4mm}
                \node (input) [draw, minimum width=\numCols*\cellSize, minimum height=\numRows*\cellSize] {
                          \begin{tikzpicture}
                                \foreach \row in {0,...,\numRows} {
                                    \draw (0, \row*\cellSize) -- (\numCols*\cellSize, \row*\cellSize);
                                }
                                
                                \foreach \col in {0,...,\numCols} {
                                    \draw (\col*\cellSize, 0) -- (\col*\cellSize, \numRows*\cellSize);
                                }
                            \end{tikzpicture}
                        };
                \node [above = 0.05cm of input]{Coarse Grid};
                \node (UNet) [draw, right=1cm of input, minimum width=7mm, minimum height=4mm] {UNet \tikzmark{A}};
                \node (Upsampler) [draw, right=1cm of UNet] {Upsampler}; 
                \def\numRows{10}
                \def\numCols{10}
                
                \def\cellSize{2mm}
                \node (output) [draw,, right=1cm of Upsampler, minimum width=\numCols*\cellSize, minimum height=\numRows*\cellSize] {
                          \begin{tikzpicture}
                                \foreach \row in {0,...,\numRows} {
                                    \draw (0, \row*\cellSize) -- (\numCols*\cellSize, \row*\cellSize);
                                }
                                
                                \foreach \col in {0,...,\numCols} {
                                    \draw (\col*\cellSize, 0) -- (\col*\cellSize, \numRows*\cellSize);
                                }
                            \end{tikzpicture}
                        };
                \node [above = 0.05cm of output]{Fine Grid};
                
                \node (Total_loss) [draw,right=1cm of output, minimum width=4cm, minimum height=2.1cm] {};
                \node (data) [draw, anchor=center,font=\scriptsize] at ([yshift=-0.5cm]Total_loss.north) {\begin{tabular}{c}Data Loss\\$u(x)-\hat{u}(x,\mathbf{W})$\end{tabular}};
                \draw [draw] (data.south)+(0cm,-0.15cm) node {+};
                \node (physics) [draw, anchor=center,font=\scriptsize] at ([yshift=0.5cm]Total_loss.south) {\begin{tabular}{c}Physics Loss \\ $N(x,\mathbf{W},u_{x},u_{xx},..)$\end{tabular}};
                \node [above=0.05cm of Total_loss] {Total loss};
                \node (Net)[draw, dashed, fit=(UNet) (Upsampler), inner sep=10pt] {};
                
                \draw[-{Latex[length=2mm,width=2.5mm]}, line width=2pt, shorten >=2pt, shorten <=2pt] (input.east) -- (UNet);
                \draw[-{Latex[length=2mm,width=2.5mm]}, line width=2pt, shorten >=2pt, shorten <=2pt] (UNet.east) -- (Upsampler);
                \draw[-{Latex[length=2mm,width=2.5mm]}, line width=2pt, shorten >=2pt, shorten <=2pt] (Upsampler.east) -- (output);
                \draw[-{Latex[length=2mm,width=2.5mm]}, line width=2pt, shorten >=2pt, shorten <=2pt] (output.east) -- (Total_loss);
                \draw[densely dotted,line width=1pt, -{Latex[length=2mm,width=2.5mm]}] (Total_loss.south) -- ([yshift=-0.325cm]Total_loss.south) -- ([yshift=-0.75cm]Net.south) -- (Net);
                \path ([xshift =-0.5cm,yshift=-2.15cm]Total_loss.south) -- node[midway, below, color=black, font=\small] {Gradient Update} ([yshift=1cm]Net.south);
                  
        \end{tikzpicture}
 n    }
    \vspace{-1.4cm}
    \caption{PIUNet Architecture}
    \label{fig:PIUNet}
\end{figure}

\begin{figure}[htbp]
    \centering
    \resizebox{\textwidth}{!}{
        \begin{tikzpicture}[node distance=0.3cm, every node/.style={rounded corners=3pt},remember picture]
            \node (Dconv1) [draw=orange!60, very thick, fill=orange!30, minimum width=1.5cm, minimum height=1.7cm] {};
            \node (conv1) [fill=orange!60, minimum width=0.05cm, minimum height=0.1cm, anchor=center] at ([xshift=-0.35cm]Dconv1.center) {\rotatebox[origin=c]{90}{($n \times$64)}};
            \node (conv2) [fill=orange!60, minimum width=0.05cm, minimum height=0.1cm, anchor=center] at ([xshift=0.35cm]Dconv1.center) {\rotatebox[origin=c]{90}{(64$\times$64)}};
            \node [below=0.05cm of Dconv1] {\textbf{D-Conv2D}};
            \node (mesh1) [draw=orange!60, rounded corners=0pt, very thick, fill=orange!30, right=of Dconv1, yshift=0.15cm, minimum width=1.2cm, minimum height=1.2cm] {};
            \node (mesh2) [draw=orange!60, rounded corners=0pt, very thick, fill=orange!30, right=0.4cm of Dconv1.south east,yshift=0.9cm, minimum width=1.2cm, minimum height=1.2cm] {};
            \node (mesh3) [draw=orange!60, rounded corners=0pt, very thick, fill=orange!30, right=0.5cm of Dconv1.south east,yshift=0.8cm, minimum width=1.2cm, minimum height=1.2cm] {};
            \node (pool1) [draw=orange!60,very thick, fill=orange!30, right=of mesh2] {\rotatebox[origin=c]{90}{MaxPool}};
    
            \node (Dconv2) [draw=orange!60, very thick, fill=orange!30, below=1cm of pool1, minimum width=1.5cm, minimum height=1.85cm] {};
            \node (conv1) [fill=orange!60, minimum width=0.05cm, minimum height=0.1cm, anchor=center] at ([xshift=-0.35cm]Dconv2.center) {\rotatebox[origin=c]{90}{(64$\times$128)}};
            \node (conv2) [fill=orange!60, minimum width=0.05cm, minimum height=0.1cm, anchor=center] at ([xshift=0.35cm]Dconv2.center) {\rotatebox[origin=c]{90}{(128$\times$128)}};
            \node [below=0.05cm of Dconv2] {\textbf{D-Conv2D}};
            \node (mesh3) [draw=orange!60, rounded corners=0pt, very thick, fill=orange!30, right=of Dconv2, yshift=0.15cm, minimum width=0.9cm, minimum height=0.9cm] {};
            \node (mesh4) [draw=orange!60, rounded corners=0pt, very thick, fill=orange!30, right=0.4cm of Dconv2.south east,yshift=1cm, minimum width=0.9cm, minimum height=0.9cm] {};
            \node (mesh5) [draw=orange!60, rounded corners=0pt, very thick, fill=orange!30, right=0.5cm of Dconv2.south east,yshift=0.9cm, minimum width=0.9cm, minimum height=0.9cm] {};
            \node (pool2) [draw=orange!60,very thick, fill=orange!30, right=of mesh4] {\rotatebox[origin=c]{90}{MaxPool}};
    
            \node (Dconv3) [draw=orange!60, very thick, fill=orange!30, below=1cm of pool2, minimum width=1.5cm, minimum height=1.85cm] {};
            \node (conv1) [fill=orange!60, minimum width=0.05cm, minimum height=0.1cm, anchor=center] at ([xshift=-0.35cm]Dconv3.center) {\rotatebox[origin=c]{90}{(128$\times$256)}};
            \node (conv2) [fill=orange!60, minimum width=0.05cm, minimum height=0.1cm, anchor=center] at ([xshift=0.35cm]Dconv3.center) {\rotatebox[origin=c]{90}{(256$\times$256)}};
            \node [below=0.05cm of Dconv3] {\textbf{D-Conv2D}};
            \node (mesh6) [draw=orange!60, rounded corners=0pt, very thick, fill=orange!30, right=of Dconv3, yshift=0.15cm, minimum width=0.75cm, minimum height=0.75cm] {};
            \node (mesh7) [draw=orange!60, rounded corners=0pt, very thick, fill=orange!30, right=0.4cm of Dconv3.south east,yshift=1cm, minimum width=0.75cm, minimum height=0.75cm] {};
            \node (mesh8) [draw=orange!60, rounded corners=0pt, very thick, fill=orange!30, right=0.5cm of Dconv3.south east,yshift=0.9cm, minimum width=0.75cm, minimum height=0.75cm] {};
            \node (pool3) [draw=orange!60,very thick, fill=orange!30, right=of mesh7] {\rotatebox[origin=c]{90}{MaxPool}};
    
            \node (Dconv4) [draw=orange!60, very thick, fill=orange!30, below=1cm of pool3, minimum width=1.5cm, minimum height=1.85cm] {};
            \node (conv1) [fill=orange!60, minimum width=0.05cm, minimum height=0.1cm, anchor=center] at ([xshift=-0.35cm]Dconv4.center) {\rotatebox[origin=c]{90}{(256$\times$512)}};
            \node (conv2) [fill=orange!60, minimum width=0.05cm, minimum height=0.1cm, anchor=center] at ([xshift=0.35cm]Dconv4.center) {\rotatebox[origin=c]{90}{(512$\times$512)}};
            \node [below=0.05cm of Dconv4] {\textbf{D-Conv2D}};
            \node (mesh9) [draw=orange!60, rounded corners=0pt, very thick, fill=orange!30, right=of Dconv4, yshift=0.15cm, minimum width=0.5cm, minimum height=0.5cm] {};
            \node (mesh10) [draw=orange!60, rounded corners=0pt, very thick, fill=orange!30, right=0.4cm of Dconv4.south east,yshift=1cm, minimum width=0.5cm, minimum height=0.5cm] {};
            \node (mesh11) [draw=orange!60, rounded corners=0pt, very thick, fill=orange!30, right=0.5cm of Dconv4.south east,yshift=0.9cm, minimum width=0.5cm, minimum height=0.5cm] {};
            \node (pool4) [draw=orange!60,very thick, fill=orange!30, right=of mesh10] {\rotatebox[origin=c]{90}{MaxPool}};

            \node (Dconv5) [draw=orange!60, very thick, fill=orange!30, below=1cm of pool4, minimum width=1.5cm, minimum height=2.3cm] {};
            \node (conv1) [fill=orange!60, minimum width=0.05cm, minimum height=0.12cm, anchor=center] at ([xshift=-0.35cm]Dconv5.center) {\rotatebox[origin=c]{90}{(512$\times$1024)}};
            \node (conv2) [fill=orange!60, minimum width=0.05cm, minimum height=0.12cm, anchor=center] at ([xshift=0.35cm]Dconv5.center) {\rotatebox[origin=c]{90}{(1024$\times$1024)}};
            \node [below=0.05cm of Dconv5] {\textbf{D-Conv2D}};
            \node (mesh12) [draw=orange!60, rounded corners=0pt, very thick, fill=orange!30, right=of Dconv5, yshift=0.15cm, minimum width=0.3cm, minimum height=0.3cm] {};
            \node (mesh13) [draw=orange!60, rounded corners=0pt, very thick, fill=orange!30, right=0.4cm of Dconv5.south east,yshift=1.2cm, minimum width=0.3cm, minimum height=0.3cm] {};
            \node (mesh14) [draw=orange!60, rounded corners=0pt, very thick, fill=orange!30, right=0.5cm of Dconv5.south east,yshift=1.1cm, minimum width=0.3cm, minimum height=0.3cm] {};
            \node (pool5) [draw=orange!60,very thick, fill=orange!30, right=of mesh13] {\rotatebox[origin=c]{90}{MaxPool}};

            \node (upconv4) [draw=violet!60,very thick, fill=violet!30, above=1cm of pool5] {\rotatebox[origin=c]{90}{(1024$\times$512)}};
            \node [left=0.02cm of upconv4] {\rotatebox[origin=c]{90}{\textbf{UpConv}}};
            \node (mesh15) [draw=violet!60, rounded corners=0pt, very thick, fill=violet!30, right=of upconv4, yshift=-0.5cm, minimum width=0.5cm, minimum height=0.5cm] {};
            \node (mesh16) [draw=violet!60, rounded corners=0pt, very thick, fill=violet!30, right=0.4cm of upconv4.south east,yshift=0.4cm, minimum width=0.5cm, minimum height=0.5cm] {};
            \node (mesh17) [draw=violet!60, rounded corners=0pt, very thick, fill=violet!30, right=0.5cm of upconv4.south east,yshift=0.3cm, minimum width=0.5cm, minimum height=0.5cm] {};
            \draw [draw=black, very thick] (mesh16.north)+(0cm,0.4cm) circle [radius=0.2cm] node {+};
            \node (mesh18) [draw=orange!60, rounded corners=0pt, very thick, fill=orange!30, right=of upconv4, above=0.8cm of mesh15, minimum width=0.5cm, minimum height=0.5cm] {};
            \node (mesh19) [draw=orange!60, rounded corners=0pt, very thick, fill=orange!30, right=0.4cm of upconv4.south east, above=0.8cm of mesh16, minimum width=0.5cm, minimum height=0.5cm] {};
            \node (mesh20) [draw=orange!60, rounded corners=0pt, very thick, fill=orange!30, right=0.5cm of upconv4.south east, above=0.8cm of mesh17, minimum width=0.5cm, minimum height=0.5cm] {};
            \node (Dconv6) [draw=violet!60, very thick, fill=violet!30, right=1.25cm of upconv4, minimum width=1.5cm, minimum height=2.3cm] {};
            \node (conv1) [fill=violet!60, minimum width=0.05cm, minimum height=0.12cm, anchor=center] at ([xshift=-0.35cm]Dconv6.center) {\rotatebox[origin=c]{90}{(1024$\times$512)}};
            \node (conv2) [fill=violet!60, minimum width=0.05cm, minimum height=0.12cm, anchor=center] at ([xshift=0.35cm]Dconv6.center) {\rotatebox[origin=c]{90}{(512$\times$512)}};
            \node [below=0.05cm of Dconv6] {\textbf{D-Conv2D}};

            \node (upconv3) [draw=violet!60,very thick, fill=violet!30, above=1cm of Dconv6] {\rotatebox[origin=c]{90}{(512$\times$256)}};
            \node [left=0.02cm of upconv3] {\rotatebox[origin=c]{90}{\textbf{UpConv}}};
            \node (mesh21) [draw=violet!60, rounded corners=0pt, very thick, fill=violet!30, right=of upconv3, yshift=-0.5cm, minimum width=0.75cm, minimum height=0.75cm] {};
            \node (mesh22) [draw=violet!60, rounded corners=0pt, very thick, fill=violet!30, right=0.4cm of upconv3.south east,yshift=0.3cm, minimum width=0.75cm, minimum height=0.75cm] {};
            \node (mesh23) [draw=violet!60, rounded corners=0pt, very thick, fill=violet!30, right=0.5cm of upconv3.south east,yshift=0.2cm, minimum width=0.75cm, minimum height=0.75cm] {};
            \draw [draw=black, very thick] (mesh22.north)+(0cm,0.4cm) circle [radius=0.2cm] node {+};
            \node (mesh24) [draw=orange!60, rounded corners=0pt, very thick, fill=orange!30, right=of upconv3, above=0.87cm of mesh21, minimum width=0.75cm, minimum height=0.75cm] {};
            \node (mesh25) [draw=orange!60, rounded corners=0pt, very thick, fill=orange!30, right=0.4cm of upconv3.south east, above=0.87cm of mesh22, minimum width=0.75cm, minimum height=0.75cm] {};
            \node (mesh26) [draw=orange!60, rounded corners=0pt, very thick, fill=orange!30, right=0.5cm of upconv3.south east, above=0.87cm of mesh23, minimum width=0.75cm, minimum height=0.75cm] {};
            \node (Dconv7) [draw=violet!60, very thick, fill=violet!30, right=1.5cm of upconv3, minimum width=1.5cm, minimum height=1.85cm] {};
            \node (conv1) [fill=violet!60, minimum width=0.05cm, minimum height=0.12cm, anchor=center] at ([xshift=-0.35cm]Dconv7.center) {\rotatebox[origin=c]{90}{(512$\times$256)}};
            \node (conv2) [fill=violet!60, minimum width=0.05cm, minimum height=0.12cm, anchor=center] at ([xshift=0.35cm]Dconv7.center) {\rotatebox[origin=c]{90}{(256$\times$256)}};
            \node [below=0.05cm of Dconv7] {\textbf{D-Conv2D}};

            \node (upconv2) [draw=violet!60,very thick, fill=violet!30, above=1cm of Dconv7] {\rotatebox[origin=c]{90}{(256$\times$128)}};
            \node [left=0.02cm of upconv2] {\rotatebox[origin=c]{90}{\textbf{UpConv}}};
            \node (mesh27) [draw=violet!60, rounded corners=0pt, very thick, fill=violet!30, right=of upconv2, yshift=-0.5cm, minimum width=0.9cm, minimum height=0.9cm] {};
            \node (mesh28) [draw=violet!60, rounded corners=0pt, very thick, fill=violet!30, right=0.4cm of upconv2.south east,yshift=0.3cm, minimum width=0.9cm, minimum height=0.9cm] {};
            \node (mesh29) [draw=violet!60, rounded corners=0pt, very thick, fill=violet!30, right=0.5cm of upconv2.south east,yshift=0.2cm, minimum width=0.9cm, minimum height=0.9cm] {};
            \draw [draw=black, very thick] (mesh28.north)+(0cm,0.4cm) circle [radius=0.2cm] node {+};
            \node (mesh30) [draw=orange!60, rounded corners=0pt, very thick, fill=orange!30, right=of upconv2, above=0.87cm of mesh27, minimum width=0.9cm, minimum height=0.9cm] {};
            \node (mesh31) [draw=orange!60, rounded corners=0pt, very thick, fill=orange!30, right=0.4cm of upconv2.south east, above=0.87cm of mesh28, minimum width=0.9cm, minimum height=0.9cm] {};
            \node (mesh32) [draw=orange!60, rounded corners=0pt, very thick, fill=orange!30, right=0.5cm of upconv2.south east, above=0.87cm of mesh29, minimum width=0.9cm, minimum height=0.9cm] {};
            \node (Dconv8) [draw=violet!60, very thick, fill=violet!30, right=1.7cm of upconv2, minimum width=1.5cm, minimum height=1.85cm] {};
            \node (conv1) [fill=violet!60, minimum width=0.05cm, minimum height=0.12cm, anchor=center] at ([xshift=-0.35cm]Dconv8.center) {\rotatebox[origin=c]{90}{(256$\times$128)}};
            \node (conv2) [fill=violet!60, minimum width=0.05cm, minimum height=0.12cm, anchor=center] at ([xshift=0.35cm]Dconv8.center) {\rotatebox[origin=c]{90}{(128$\times$128)}};
            \node [below=0.05cm of Dconv8] {\textbf{D-Conv2D}};

            \node (upconv1) [draw=violet!60,very thick, fill=violet!30, above=1cm of Dconv8] {\rotatebox[origin=c]{90}{(128$\times$64)}};
            \node [left=0.02cm of upconv1] {\rotatebox[origin=c]{90}{\textbf{UpConv}}};
            \node (mesh33) [draw=violet!60, rounded corners=0pt, very thick, fill=violet!30, right=of upconv1, yshift=-0.45cm, minimum width=1.2cm, minimum height=1.2cm] {};
            \node (mesh34) [draw=violet!60, rounded corners=0pt, very thick, fill=violet!30, right=0.4cm of upconv1.south east,yshift=0.3cm, minimum width=1.2cm, minimum height=1.2cm] {};
            \node (mesh35) [draw=violet!60, rounded corners=0pt, very thick, fill=violet!30, right=0.5cm of upconv1.south east,yshift=0.2cm, minimum width=1.2cm, minimum height=1.2cm] {};
            \draw [draw=black, very thick] (mesh34.north)+(0cm,0.4cm) circle [radius=0.2cm] node {+};
            \node (mesh36) [draw=orange!60, rounded corners=0pt, very thick, fill=orange!30, right=of upconv1, above=0.87cm of mesh33, minimum width=1.2cm, minimum height=1.2cm] {};
            \node (mesh37) [draw=orange!60, rounded corners=0pt, very thick, fill=orange!30, right=0.4cm of upconv1.south east, above=0.87cm of mesh34, minimum width=1.2cm, minimum height=1.2cm] {};
            \node (mesh38) [draw=orange!60, rounded corners=0pt, very thick, fill=orange!30, right=0.5cm of upconv1.south east, above=0.87cm of mesh35, minimum width=1.2cm, minimum height=1.2cm] {};
            \node (Dconv9) [draw=violet!60, very thick, fill=violet!30, right=1.95cm of upconv1,minimum width=1.5cm, minimum height=1.85cm] {};
            \node (conv1) [fill=violet!60, minimum width=0.05cm, minimum height=0.12cm, anchor=center] at ([xshift=-0.35cm]Dconv9.center) {\rotatebox[origin=c]{90}{(128$\times$64)}};
            \node (conv2) [fill=violet!60, minimum width=0.05cm, minimum height=0.12cm, anchor=center] at ([xshift=0.35cm]Dconv9.center) {\rotatebox[origin=c]{90}{(64$\times$64)}};
            \node [above=0.05cm of Dconv9] {\textbf{D-Conv2D}};
    
            \node (Outconv) [draw=violet!60,very thick, fill=violet!30, below=1.5cm of Dconv9] {(64$\times n$)};
            \node [above=0.05cm of Outconv] {\textbf{OutConv}};

            \draw[-{Latex[length=2mm,width=2.5mm]}, line width=2pt, shorten >=2pt, shorten <=2pt] (pool1.south) -- (Dconv2);
            \draw[-{Latex[length=2mm,width=2.5mm]}, line width=2pt, shorten >=2pt, shorten <=2pt] (pool2.south) -- (Dconv3);
            \draw[-{Latex[length=2mm,width=2.5mm]}, line width=2pt, shorten >=2pt, shorten <=2pt] (pool3.south) -- (Dconv4);
            \draw[-{Latex[length=2mm,width=2.5mm]}, line width=2pt, shorten >=2pt, shorten <=2pt] (pool4.south) -- (Dconv5);
            \draw[-{Latex[length=2mm,width=2.5mm]}, line width=2pt, shorten >=2pt, shorten <=2pt] (pool5.north) -- (upconv4);
            \draw[-{Latex[length=2mm,width=2.5mm]}, line width=2pt, shorten >=2pt, shorten <=2pt] (Dconv6.north) -- (upconv3);
            \draw[-{Latex[length=2mm,width=2.5mm]}, line width=2pt, shorten >=2pt, shorten <=2pt] (Dconv7.north) -- (upconv2);
            \draw[-{Latex[length=2mm,width=2.5mm]}, line width=2pt, shorten >=2pt, shorten <=2pt] (Dconv8.north) -- (upconv1);
            \draw[-{Latex[length=2mm,width=2.5mm]}, line width=2pt, shorten >=14pt, shorten <=2pt] (Dconv9.south) -- (Outconv);

            \draw[-{Latex[length=5mm,width=2.5mm]}, line width=2pt, purple, shorten >=5pt, shorten <=10pt] (mesh1.north) to [bend left=15] ([yshift=-0.5cm]mesh36.west);
            \path ([xshift = 21cm,yshift=3.5cm]mesh1.north) -- node[midway, above, color=black] {\textbf{Skip Connection}} ([yshift=1cm]mesh1.west);
            \draw[-{Latex[length=5mm,width=2.5mm]}, line width=2pt, purple, shorten >=5pt, shorten <=5pt] ([yshift=0.2cm]mesh4.north) to [bend left=15] (mesh30.west);
            \path ([xshift =2cm,yshift=1.5cm]mesh4.north) -- node[midway, above, color=black] {\textbf{Skip Connection}} ([yshift=1cm]mesh30.west);
            \draw[-{Latex[length=5mm,width=2.5mm]}, line width=2pt, purple, shorten >=5pt, shorten <=5pt] ([yshift=0.2cm]mesh7.north) to [bend left=17] (mesh24.west);
            \path ([xshift =0.75cm,yshift=0.75cm]mesh7.north) -- node[midway, above, color=black] {\textbf{Skip Connection}} ([yshift=1cm]mesh24.west);
            \draw[-{Latex[length=5mm,width=2.5mm]}, line width=2pt, purple, shorten >=5pt, shorten <=5pt] ([yshift=0.2cm]mesh10.north) to [bend left=45] (mesh18.north);
            \path ([xshift =0.5cm,yshift=1cm]mesh10.north) -- node[midway, above, color=black] {\textbf{Skip Connection}} ([yshift=1cm]mesh18.north);

          \end{tikzpicture}

    }
    \caption{UNet Architecture}
    \label{fig:UNet}
\end{figure}
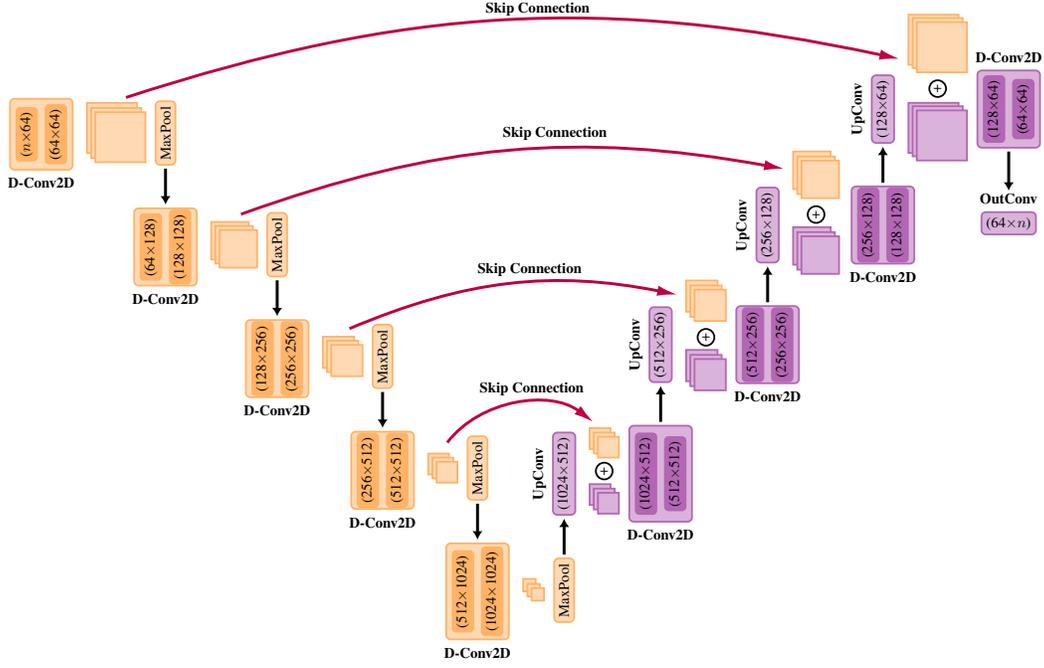

The architecture of our Physics Informed UNet (PIUNet) for upscaling coarse-grid fluid flow simulation data is depicted in Figure \ref{fig:PIUNet}. PIUNet bridges the gap between low-resolution coarse mesh data and high-resolution fine mesh data, enabling accurate fluid flow behavior predictions. The PIUNet begins with four contracting convolution layers. Each layer comprises a 3x3 double unpadded convolution operation, followed by ReLU activation, and subsequently a 2x2 max-pooling layer. These layers progressively downsample the input coarse grid data, allowing the network to capture increasingly abstract and high-level features. It is followed by a bottleneck layer capturing high-level features while preserving spatial information. Following the bottleneck layer, four expansion convolution layers are employed. Each layer consists of a deconvolution operation, a skip connection, and another 3x3 double unpadded convolution operation, followed by ReLU activation. The skip connection guides the upsampling process, retaining essential spatial information. After the expansion layers, a 1x1 convolution layer reconstructs the output to match the dimension of the coarse mesh data. A 1x1 convolution layer precedes the bilinear upsampler, which leverages the learned representations from the network to upsample the data to the fine-mesh dimension effectively. The detailed UNet architecture is visually represented in Figure \ref{fig:UNet}. In the training process, we employ a customized loss function that combines both data loss and physics loss. This loss function encourages and updates the gradient of the network to learn not only from the data but also from the underlying physics of the system. 

\subsection{Loss Function:}
In the training process, we utilize a customized loss function that combines data and physics components. The data loss ($\mathcal{L}_{data}$) is computed as the Mean Square Error (MSE) between fine mesh and predicted fine mesh data, while physics loss ($\mathcal{L}_{physics}$) measures the MSE of the convective and diffusive terms from the governing equations. The total loss function is given by:
\begin{equation}
    \hspace{-4cm}\begin{aligned}
     &\text{Total Loss Function:}\hspace{3cm} \mathcal{L}_{total} =  \mathcal{L}_{data} + \mathcal{L}_{physics}
    \end{aligned}
\end{equation}
The physics loss is tailored for different datasets with distinct weights, the details of these weights and the equation of the physics loss for each respective dataset can be found in the Appendix \ref{sec:Loss_function}.

\section{Results and Discussion}
In this section, we present the outcomes of our experiments, focusing on the performance evaluation of the PIUNet across various datasets as previously mentioned. Detailed results are summarized in Table \ref{tab:Result}, providing comprehensive insights into our findings. We employed two widely recognized interpolation techniques, namely Bi-linear Interpolation and Bi-cubic Interpolation as baseline methods to compare the super-resolution capabilities for upscaling coarse mesh data to fine mesh data features. Initially, we utilized a vanilla UNet-based super-resolution approach. As we sought to improve the results, we integrated the physics loss, resulting in the development of PIUNet. We have applied this to three CFD problems, phenomena represented by Burger's Equation, fouling in industrial heat exchangers, and methane combustion.

\textbf{Burger's Equation:}\cite{mathias2022augmenting} Comparing the UNet-based super-resolution approach with baseline interpolations, we observe significant improvements. In terms of RMSE, we achieve approximately 18x and 14x enhancements for $U_x$ and $U_y$ in the $x$ and $y$ directions, respectively. Additionally, MAE demonstrates improvements of roughly 30x for $U_x$ and 32x for $U_y$. The introduction of physics into UNet, i.e., the Physics-Informed UNet, further enhances results by reducing RMSE by approximately 27\% for $U_x$ and 27\% for $U_y$.

\textbf{Industrial Heat Exchanger:}\cite{majumdar2022real} This problem involves six features, namely fluid temperatures $T_1$,$T_2$ and $T_3$, and matrix temperatures $T_{m_1}$,$T_{m_2}$,$T_{m_3}$. Similar to Burger's equation, the UNet-based super-resolution method outperforms baseline interpolations. The Physics-Informed UNet further enhances results, achieving a 36.8\% reduction in RMSE for $T_{1}$, 30.7\% for $T_{2}$ and $T_3$, 35.04\% for $T_{m_1}$, and 20.06\% for $T_{m_2}$,$T_{m_3}$. MAE also witnesses improvements of approximately 37\% for $T_{1}$, 50\% for $T_{2}$,$T_{3}$, 39.2\% for $T_{m_1}$, and 27.48\% for $T_{m_2}$ and $T_{m_3}$.

\textbf{Methane Combustion:}\cite{yang2019reactingfoam} This problem involves seven outputs, including adiabatic flame temperature ($T_{adia}$), $x$-direction velocity, $y$-direction velocity, and mass fractions of species $\mathrm{CH}_4$, $\mathrm{O}_2$, $\mathrm{H}_2\mathrm{O}$, and $\mathrm{CO}_2$. Notably, the rise in adiabatic flame temperature and the depletion of the $\mathrm{CH}_4$ mass fraction significantly influence combustion. The Physics-Informed UNet-based super-resolution method outperforms baseline interpolation methods, with RMSE improvements of 73\% in $T_{adia}$ and 20.47\% in the mass fraction of $\mathrm{CH}_4$. For other features such as velocities and mass fractions of $\mathrm{O}_2$, $\mathrm{H}_2\mathrm{O}$, and $\mathrm{CO}_2$, results align closely with baseline interpolation due to minimal differences between coarse and fine mesh data. 

\begin{table}[htbp]
\centering
\caption{Performance Comparison of PIUNet vs. Baseline Interpolations on test data using RMSE, MAE, and R$^2$ Metrics for Diverse Datasets(Best score are in bold)}
\label{tab:Result}
\resizebox{\textwidth}{!}{%
\begin{tabular}{@{}c|ccccc|c|ccccc@{}}
\toprule
\textbf{}                                & \textbf{Features}                                                                  & \textbf{Algorithm} & \textbf{RMSE}   & \textbf{MAE}     & \textbf{R$^{2}$}     & \multicolumn{1}{l|}{}                         & \textbf{Features}            & \textbf{Algorithm} & \textbf{RMSE}   & \textbf{MAE}    & \textbf{R$^{2}$}     \\ \midrule
\multirow{9}{*}{\rotatebox[origin=c]{90}{\textbf{Burger's Eq}}}     & \multicolumn{5}{c|}{51$\times$51 $\rightarrow$ 401$\times$401}                                                                                                                         & \multirow{30}{*}{\rotatebox[origin=c]{90}{\textbf{Methane Combustion}}} & \multicolumn{5}{c}{50$\times$20 $\rightarrow$ 500$\times$100}                                                                   \\ \cmidrule(lr){2-6} \cmidrule(l){8-12} 
                                         & \multirow{4}{*}{\begin{tabular}[c]{@{}c@{}}X-Velocity\\ ($U_{x}$)\end{tabular}}                                                        & Bilinear           & 0.4927          & 0.3858           & -0.3368         &                                               & \multirow{4}{*}{\begin{tabular}[c]{@{}c@{}}Temperature\\ ($T_{adia})$\end{tabular}} & Bilinear           & 78.317          & 32.444          & 0.9776          \\
                                         &                                                                                    & Bicubic            & 0.5133          & 0.3927           & -0.4513         &                                               &                              & Bicubic            & 77.374          & 31.486          & 0.9779          \\
                                         &                                                                                    & UNet               & 0.0283          & 0.0126           & \textbf{0.9955} &                                               &                              & UNet               & 30.718          & 13.224          & 0.9963          \\
                                         &                                                                                    & \textbf{PIUNet}    & \textbf{0.0207} & \textbf{0.0062}  & 0.9948          &                                               &                              & \textbf{PIUNet}    & \textbf{20.954} & \textbf{10.385} & \textbf{0.9984} \\ \cmidrule(lr){2-6} \cmidrule(l){8-12} 
                                         & \multirow{4}{*}{\begin{tabular}[c]{@{}c@{}}Y-Velocity\\ ($U_{y}$)\end{tabular}}                                                        & Bilinear           & 0.3517          & 0.2602           & -0.4616         &                                               & \multirow{4}{*}{\begin{tabular}[c]{@{}c@{}}X-Velocity\\ ($U_{x}$)\end{tabular}}  & Bilinear           & 0.0289          & 0.0122          & 0.9843          \\
                                         &                                                                                    & Bicubic            & 0.3804          & 0.2752           & -0.7106         &                                               &                              & Bicubic            & 0.0290          & 0.0121          & 0.9842          \\
                                         &                                                                                    & UNet               & 0.0247          & 0.00817          & 0.9927          &                                               &                              & UNet               & 0.0296          & 0.0177          & 0.9839          \\
                                         &                                                                                    & \textbf{PIUNet}    & \textbf{0.0225} & \textbf{0.00289} & \textbf{0.9965} &                                               &                              & \textbf{PIUNet}    & \textbf{0.0286} & \textbf{0.0164} & \textbf{0.9862} \\ \cmidrule(r){1-6} \cmidrule(l){8-12} 
\multirow{21}{*}{\rotatebox[origin=c]{90}{\textbf{Industrial Heat Exchanger}}} & \multicolumn{5}{c|}{30$\times$30 $\rightarrow$ 480$\times$480}                                                                                                                         &                                               & \multirow{4}{*}{\begin{tabular}[c]{@{}c@{}}Y-Velocity\\ ($U_{y}$)\end{tabular}}  & Bilinear           & 0.0325          & \textbf{0.0118} & 0.9819          \\ \cmidrule(lr){2-6} 
                                         & \multirow{4}{*}{\begin{tabular}[c]{@{}c@{}}Gas flow \\ Temperature \\ ($T_1$)\end{tabular}}   & Bilinear           & 3.1859          & 2.4508           & 0.9971          &                                               &                              & Bicubic            & 0.0332          & 0.0125          & 0.9815          \\
                                         &                                                                                    & Bicubic            & 3.2192          & 2.4635           & 0.9971          &                                               &                              & UNet               & 0.0324          & 0.0152          & 0.9830          \\
                                         &                                                                                    & UNet               & 2.5920          & 2.0306           & 0.9940          &                                               &                              & \textbf{PIUNet}    & \textbf{0.0324} & 0.0158          & \textbf{0.9835} \\ \cmidrule(l){8-12} 
                                         &                                                                                    & \textbf{PIUNet}    & \textbf{2.0112} & \textbf{1.5386}  & \textbf{0.9985} &                                               & \multicolumn{5}{c}{Mass Fractions}                                                                      \\ \cmidrule(lr){2-6} \cmidrule(l){8-12} 
                                         & \multirow{4}{*}{\begin{tabular}[c]{@{}c@{}}Air flow \\ Temperature \\ ($T_2$ and $T_3$)\end{tabular}}   & Bilinear           & 4.0527          & 3.4445           & 0.9980          &                                               & \multirow{4}{*}{$\mathrm{CH}_4$}         & Bilinear           & 0.0356          & 0.0104          & 0.9920          \\
                                         &                                                                                    & Bicubic            & 4.0008          & 3.4324           & 0.9980          &                                               &                              & Bicubic            & 0.0357          & 0.0105          & 0.9920          \\
                                         &                                                                                    & UNet               & 3.7728          & 2.8390           & 0.9981          &                                               &                              & UNet               & 0.0321          & 0.0140          & 0.9938          \\
                                         &                                                                                    & \textbf{PIUNet}    & \textbf{2.8084} & \textbf{1.7214}  & \textbf{0.9988} &                                               &                              & \textbf{PIUNet}    & \textbf{0.0283} & \textbf{0.0138} & \textbf{0.9954} \\ \cmidrule(lr){2-6} \cmidrule(l){8-12} 
                                         & \multirow{4}{*}{\begin{tabular}[c]{@{}c@{}}Gas Matrix \\ Temperature \\ ($T_{m_1}$)\end{tabular}} & Bilinear           & 3.0919          & 2.6329           & 0.9981          &                                               & \multirow{4}{*}{$\mathrm{O}_2$}          & Bilinear           & 0.0105          & 0.0028          & 0.9889          \\
                                         &                                                                                    & Bicubic            & 3.1228          & 2.6244           & 0.9981          &                                               &                              & Bicubic            & \textbf{0.0104} & \textbf{0.0027} & \textbf{0.9891} \\
                                         &                                                                                    & UNet               & 2.5787          & 2.0840           & 0.9979          &                                               &                              & UNet               & 0.0116          & 0.0059          & 0.9870          \\
                                         &                                                                                    & \textbf{PIUNet}    & \textbf{2.0084} & \textbf{1.5985}  & \textbf{0.9987} &                                               &                              & \textbf{PIUNet}    & 0.0106          & 0.0030          & 0.9888          \\ \cmidrule(lr){2-6} \cmidrule(l){8-12} 
                                         & \multirow{4}{*}{\begin{tabular}[c]{@{}c@{}}Air Matrix \\ Temperature \\ ($T_{m_2}$)\end{tabular}} & Bilinear           & 3.6404          & 3.1354           & 0.9979          &                                               & \multirow{4}{*}{$\mathrm{H}_2\mathrm{O}$}         & Bilinear           & 0.0038          & 0.0014          & 0.9893          \\
                                         &                                                                                    & Bicubic            & 3.5790          & 3.1288           & 0.9980          &                                               &                              & Bicubic            & \textbf{0.0037} & \textbf{0.0013} & \textbf{0.9899} \\
                                         &                                                                                    & UNet               & 4.2434          & 3.6032           & 0.9971          &                                               &                              & UNet               & 0.0076          & 0.0037          & 0.9598          \\
                                         &                                                                                    & \textbf{PIUNet}    & \textbf{2.9098} & \textbf{2.2735}  & \textbf{0.9982} &                                               &                              & \textbf{PIUNet}    & 0.0040          & 0.0020          & 0.9891          \\ \cmidrule(lr){2-6} \cmidrule(l){8-12} 
                                         & \multirow{4}{*}{\begin{tabular}[c]{@{}c@{}}Air Matrix \\ Temperature \\ ($T_{m_3}$)\end{tabular}} & Bilinear           & 3.6404          & 3.1354           & 0.9979          &                                               & \multirow{4}{*}{$\mathrm{CO}_2$}         & Bilinear           & 0.0052          & 0.0019          & 0.9844          \\
                                         &                                                                                    & Bicubic            & 3.5790          & 3.1288           & 0.9980          &                                               &                              & Bicubic            & 0.0052          & 0.0018          & \textbf{0.9849} \\
                                         &                                                                                    & UNet               & 4.2434          & 3.6032           & 0.9971          &                                               &                              & UNet               & 0.0094          & 0.0058          & 0.9564          \\
                                         &                                                                                    & \textbf{PIUNet}    & \textbf{2.9098} & \textbf{2.2735}  & \textbf{0.9982} &                                               &                              & \textbf{PIUNet}    & \textbf{0.0051} & \textbf{0.0018} & 0.9844          \\ \bottomrule
\end{tabular}%
}
\end{table}
Furthermore, as evident in Table \ref{tab:Cost}, our utilization of the Physics-Informed UNet in conjunction with the physics model has resulted in a substantial reduction in computational costs for fine mesh simulations.
\begin{table}[htbp]
\centering
\caption{Accelerated CFD Simulations: PIUNet's speedup compared to traditional techniques}
\label{tab:Cost}
\resizebox{\textwidth}{!}{%
\scriptsize
\begin{tabular}{cccc}
\toprule
\textbf{CFD problem} & \textbf{Simulation technique} & \textbf{Grid size} & \textbf{Simulation Time (s)} \\
\midrule
\multirow{3}{*}{2D Burger’s Equation} & \multirow{2}{*}{\begin{tabular}[c]{@{}c@{}}Finite Difference Method \\ (MATLAB)\end{tabular}} & 51 $\times$ 51 & 151 \\
& & 401 $\times$ 401 & 3623 \\
& FDM + PIUNet trained model & 401 $\times$ 401 & \textbf{152.764} ($\approx$\textbf{24X speed}) \\
\midrule
\multirow{3}{*}{Counterflow Methane Combustion} & \multirow{2}{*}{\begin{tabular}[c]{@{}c@{}}Finite Volume Method \\ (Open FOAM)\end{tabular}} & 50 $\times$ 20 & 74.59 \\
& & 500 $\times$ 100 & 7146.67 \\
& FDM + PIUNet & 500 $\times$ 100 & \textbf{76.795} ($\approx$\textbf{93X speed}) \\
\midrule
\multirow{3}{*}{Industrial Heat Exchanger} & \multirow{2}{*}{\begin{tabular}[c]{@{}c@{}}Finite Difference Method \\ (Python)\end{tabular}} & 30 $\times$ 30 & 6.969 \\
& & 480 $\times$ 480 & 2200.45 \\
& FDM + PIUNet & 480 $\times$ 480 & \textbf{9.369} ($\approx$\textbf{235X speed}) \\
\bottomrule
\end{tabular}%
}
\end{table}

\section{Conclusion}
In conclusion, our study highlights the effectiveness of the PIUNet in improving super-resolution tasks of upsampling the coarse grid data to fine grid data across different datasets. For Burger's equation, we achieved substantial RMSE and MAE enhancements, especially for fluid velocity in the $x$ and $y$ directions, by incorporating physics into the UNet. In the industrial heat exchanger scenario, PIUNet outperformed baseline methods significantly reducing RMSE and MAE for various temperature parameters. In the case of methane combustion, PIUNet notably improved RMSE for adiabatic flame temperature and $\mathrm{CH}_4$ mass fraction. These results demonstrate PIUNet's potential in capturing complex physical behaviors in different boundary conditions, with promising applications in the fluid dynamics of industrial systems.

\section*{Impact Statement}
The current state-of-the-art super-resolution techniques involve reconstructing high-resolution data from downsampled low-resolution data, and face challenges when applied to traditional CFD simulations. These challenges stem from the scarcity of high-resolution mesh data for training, and from the fundamental differences between downsampling high-resolution data and simulating low-resolution data using traditional numerical solvers. This distinction arises from the complex physical properties that are inherently associated with high-resolution data. Simply downsampling from high-resolution data results in retaining the majority of the governing physics, which may not accurately reflect real-world scenarios. We utilize coarse-grid data as the input to predict fine-grid results in super-resolution, seamlessly integrating it into traditional CFD models. This eliminates the need for resource-intensive fine-mesh CFD simulations, significantly reducing computation time while preserving intricate fluid behavior details. This advancement can enhance the accuracy of computational simulations, leading to improved designs and operations optimization in industries such as aerospace, energy, etc. However, it's important to acknowledge the limitation of our current work, which lies in the specific grid upsampling training. As a future endeavor, we plan to modify our model to overcome this limitation and extend its applicability to irregular and complex geometry meshes.   

\section*{Acknowledgment} 
The author extends his appreciation to TCS Research for the opportunity to undertake this project. A special acknowledgment goes to co-authors Ritam Majumdar, Vishal Jadhav, Dr.Sagar Sakhinanana, and Dr.Venkataramana Runkana for their consistent support, guidance, and invaluable insights throughout the research process. Gratitude is also extended to colleagues Shivam Gupta and Krishna Sai Sudhir Aripirala for their assistance in generating the dataset, a pivotal aspect of the project. Additionally, thanks are conveyed to the organizers and reviewers of the NeurIPS ML4PS workshop for providing a platform to disseminate our research findings. Finally, the author expresses thanks to family and friends for their encouragement and understanding throughout this academic pursuit.
{
\small
\bibliographystyle{plain}
\bibliography{reference}
}
\vspace{10cm}
\section{Appendix}
\subsection{Training and Hyperparameter details}
In our experiments, we conducted training using the PyTorch framework on a Nvidia\textregistered P100 GPU with 16GB of memory. We generated three distinct two-dimensional datasets for Burger equation simulation, Counterflow methane combustion simulation, and industrial heat exchanger simulation.. For the 2D-Burger dataset, we produced 10 boundary conditions, each comprising 200-time steps of coarse mesh data (51$\times$51 grid) paired with fine mesh data (401$\times$401 grid). In the 2D-counterflow methane combustion dataset, there were 10 boundary conditions, each spanning 50-time steps, with coarse (50$\times$20) and fine (500$\times$100) mesh data. The industrial heat exchanger dataset included 100 boundary conditions of steady state, with coarse mesh data (30$\times$30 grid) and fine mesh data (480$\times$480 grid). During training, we employed the Adam optimizer for 500 epochs, starting with a learning rate of $1e^{-3}$ and weight decay of $5e^{-4}$. The learning rate was reduced by 30\% after 5 epochs without validation improvement. We determined the total dataset size for each case by multiplying the number of boundary conditions by the number of time steps. Data was partitioned into training, validation, and testing sets using a 60/20/20 ratio for evaluation. All the distinct weights of the physics loss of the different datasets are the hyperparameters of our model.
\subsection{UNet Architecture}
\subsection{Dataset Information}
\label{sec:dataset}
\textbf{Burger's Equation} We investigate the behavior of wave phenomena in a viscous fluid flow by considering the 2D Burger's equation. This equation is a fundamental non-linear advection-diffusion problem that characterizes the dynamics of wave-like structures in a two-dimensional fluid medium. The 2D Burger's Equation is mathematically expressed as follows:
\begin{equation}
\partial_t \mathbf{u}+\mathbf{u} \cdot \nabla \mathbf{u} =\nu\left(\nabla^2 \mathbf{u}\right) 
\end{equation}
Here, the variable $\mathbf{u}$ represents the velocity field and $\nu$ represents kinematic viscosity. For our investigations, we have adopted sinusoidal initial conditions, $U_{x}(0, x, y) = \sin (2 \pi x) \sin (2 \pi y)$ and $U_{y}(0, x, y) = \sin (\pi x) \sin (\pi y)$. These initial conditions, as detailed in \cite{mathias2022augmenting}, serve as the starting point for our experiments. Specifically, we explore a range of Dirichlet boundary conditions for $\mathbf{u}$ within the intervals (0, 1)m/s to gain insights into the diverse behaviors exhibited by the system. The 2D-Burger simulation dataset was generated in a rectangular domain using MATLAB-based code\cite{mathias2022augmenting}. This code approximated spatial and time derivatives using 6th-order FDM and 4th-order Runge-Kutta schemes with a time step ($\Delta t$) of 10$^{-5}$

\textbf{Methane Combustion} We investigate the fundamental chemical reaction of methane combustion within a 2D-Laminar counter-flow configuration. This reaction is pivotal in various combustion systems, such as gas turbines and furnaces. In the counter-flow setup, the velocities of methane (the fuel) and air are oriented in opposing directions. The chemical reaction governing the combustion of methane, $\mathrm{CH}_4$, with air is as follows:
\begin{equation*}
\mathrm{CH}_4 + 2\mathrm{O}_2 \rightarrow \mathrm{CO}_2 + 2\mathrm{H}_2\mathrm{O} 
\end{equation*}
Our analysis is based on a set of governing equations that describe the key physical aspects of this combustion process: 
\begin{equation}
\hspace{-1.2cm}\begin{aligned}
    &\text{continuity eq:} \hspace{1.8cm} \partial_t \rho+\nabla \cdot(\rho \mathbf{u})=0\\
    &\text{momentum eq:} \hspace{1.6cm} \partial_t(\rho \mathbf{u})+\nabla \cdot(\rho \mathbf{u} \mathbf{u})=-\nabla p+\mu(\nabla^2 \mathbf{u}) \\
    &\text{species transport eq:} \hspace{0.8cm} \partial_t\left(\rho Y_i\right)+\nabla \cdot\left(\rho \mathbf{u} Y_i\right)=\mu\nabla^2 Y_i+\dot{R}_i\\
    &\text{energy eq:} \hspace{2.2cm}\partial_t(\rho h)+\nabla(\rho \mathbf{u} h)+\partial_t(\rho K)+\nabla(\rho \mathbf{u} K)-\partial_t p=\alpha \nabla^2 h+\dot{R}_{\text {heat}} 
\end{aligned}
\end{equation}
In these equations, $\mathbf{u}$ represents the velocity vector, $\rho$ denotes density, $p$ stands for pressure, and $\mu$ represents the dynamic viscosity of the fluid mixture. The species transport equation involves $Y_i$ (mass fraction of species $i$) and $\dot{R}_i$ (production rate of species $i$). The energy equation incorporates $h$ (internal energy), where $\alpha$, $K$, and $\dot{R}_{\text{heat}}$ correspond to thermal diffusivity, kinematic energy, and heat generation due to the chemical reaction, respectively. Considering isobaric combustion, all pressure gradient terms become zero. Our investigation includes a range of Dirichlet boundary conditions for $\mathbf{u}$ for both fuel and air within the intervals (0.1, 0.6) m/s. This exploration provides valuable insights into the diverse behaviors exhibited by the combustion system. The simulation dataset was conducted within a 2cm$\times$2cm domain by employing OpenFOAM reactingFoam solver\cite{yang2019reactingfoam} while assuming laminar flow conditions.

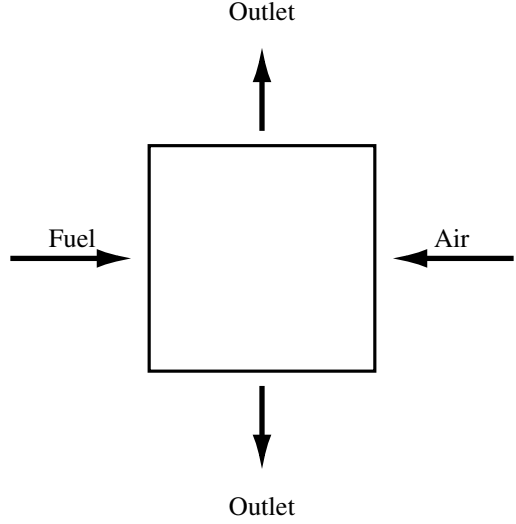
\begin{figure}
    \centering
    \begin{tikzpicture}
        \node(square)[draw, very thick, minimum width=3cm, minimum height=3cm]{};
    
        \draw[-{Latex[length=5mm,width=2.5mm]}, line width=2pt, shorten >=5pt, shorten <=5pt] ([xshift=-2cm]square.west) -- (square.west) node[midway,above] {Fuel};
    
        \draw[-{Latex[length=5mm,width=2.5mm]}, line width=2pt, shorten >=5pt, shorten <=5pt] ([xshift=2cm]square.east) -- (square.east) node[midway,above] {Air};
    
        \draw[-{Latex[length=5mm,width=2.5mm]}, line width=2pt, shorten >=5pt, shorten <=5pt] (square.north) -- ([yshift=1.5cm]square.north) node[above] {Outlet};
    
        \draw[-{Latex[length=5mm,width=2.5mm]}, line width=2pt, shorten >=5pt, shorten <=5pt] (square.south) -- ([yshift=-1.5cm]square.south) node[below] {Outlet};
    \end{tikzpicture}
    \caption{Computational Domain of a 2D-counterflow methane combustion}
    \label{fig:my_label}
\end{figure}
\begin{figure}[htbp]
    \centering
    \includegraphics[scale=0.35]{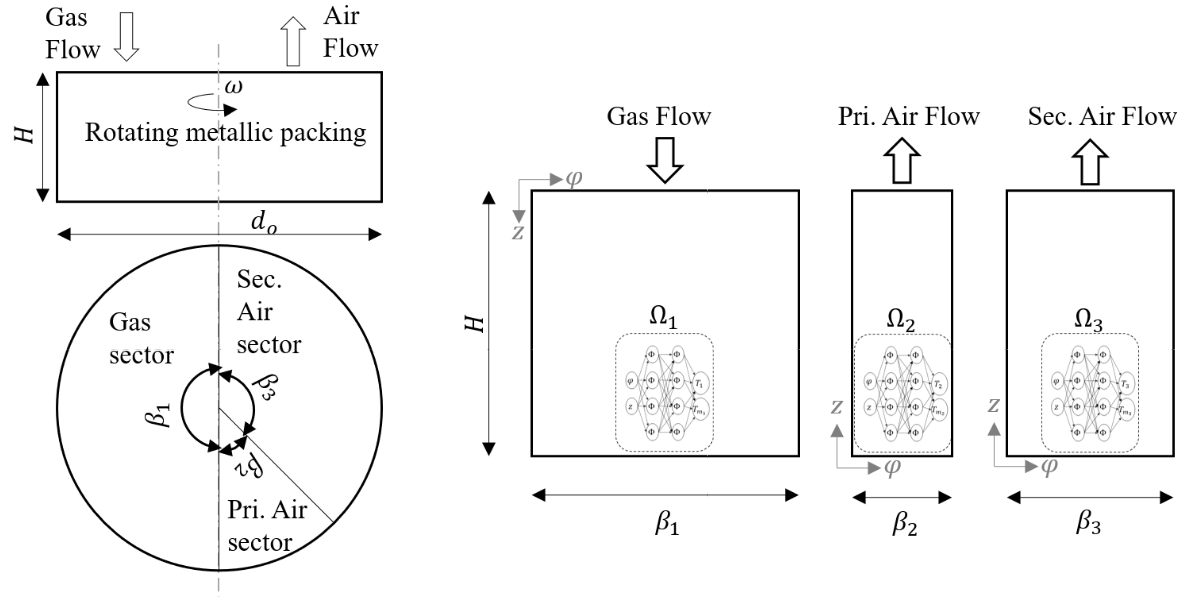}
    \caption{Computational Domain of an Industrial Heat Exchanger}
    \label{fig:Computational Domain of APH}
\end{figure}

\textbf{Industrial Heat exchanger} We explore the non-dimensional, 2D formulation of counterflow heat transfer within an industrial heat exchanger. These heat exchangers play a vital role in thermal power plants, enhancing overall thermal efficiency. Monitoring the internal temperature profiles of these heat exchangers is crucial to prevent failures stemming from intricate thermal and chemical deposition phenomena. We use Eq \ref{eq:cond} for heat conduction and Eq \ref{eq:conv} for convective heat transfer. Our system has six outputs: three fluid temperatures ($T$) and three metal temperatures ($T_m$) at specific coordinates ($\varphi$, $z$). Given that this equipment operates under mostly consistent conditions, with minimal fluctuations in flow patterns, we have opted to utilize steady-state energy transfer equations to elucidate the heat transfer phenomena. It is crucial to emphasize that, in this context, the preservation of energy takes on great significance due to the notable variations in temperature gradients that arise from the processes of convection and conduction along the longitudinal axis of the matrix.\cite{skiepko1988effect}.

\begin{equation}
\label{eq:cond}
    \partial_{\varphi} T_{m_{j}} = NTU_{m_{j}} (T_j - T_{m_j}) + Pe_{m_j}^{-1}   \partial_{z}^2 T_{m_j}
\end{equation}
\begin{equation}
\label{eq:conv}
    \partial_z T_j = {NTU}_{m_j} \left( {T}_{m_{j}} - {T}_{j}, \right) \hspace{0.2cm} j = 1,2,3
\end{equation}
 The non-dimensional boundary conditions are set by gas($T_{in,1}$), primary air($T_{in,2}$), and secondary air inlet temperatures($T_{in,3}$) is given by,
\begin{equation}
\begin{aligned}
    &{T_j}\left( \varphi, z=0 \right) = {T}_{in,j},  \hspace{0.2cm} j = 1,2,3 \\
    &T_{m_1}(\varphi = 0, z) = T_{m_3} (\varphi = 1, 1-z)\\
    &T_{m_1}(\varphi = 1, z) = T_{m_2} (\varphi = 0, 1 - z)\\
    &T_{m_2} (\varphi = 1, z) = T_{m_3} (\varphi = 0, z),\\ 
    &\partial_z {T}_{m_j}[z = 0, 1] = 0,  \hspace{0.2cm}j = 1,2,3 
\end{aligned}
\end{equation}
the above matrix boundary conditions impose continuity constraints on metal temperatures. The simulation dataset is generated in a cylindrical domain as shown in figure \ref{fig:Computational Domain of APH} using finite difference method-based solver detailed in \cite{li1983numerical}

\subsection{Physics Loss Function:}
\label{sec:Loss_function} 
In our training process, we implement a tailored loss function that combines data loss and physics loss components. The data loss ($\mathcal{L}$) is calculated as the Mean Square Error (MSE) between the fine mesh data and the predicted fine mesh data. The physics loss, on the other hand, is determined by computing the MSE of the residual of the governing equations between the fine mesh and the predicted mesh data. The total loss function is expressed as:
\begin{equation}
    \hspace{-4cm}\begin{aligned}
     &\text{Total Loss Function:}\hspace{3cm} \mathcal{L}_{total} =  \mathcal{L}_{data} + \mathcal{L}_{physics}
    \end{aligned}
\end{equation}
We have tailored the weighted physics loss for our dataset to capture the relevant physical constraints. Adjustments to this loss expression can be made to adapt to different physics scenarios within the dataset. The physics losses for the different datasets are defined here.
\begin{equation}
    \hspace{-3.2cm}\begin{aligned}
     &\text{Physics Loss:}\hspace{3cm} \mathcal{L}_{physics} =  \alpha_{conv}\mathcal{L}_{conv} + \alpha_{diff}\mathcal{L}_{diff} 
    \end{aligned}
\end{equation}
\textbf{2D-Burger Equation}
We have set the weight for the convective term $\alpha_{conv}$ is 1$e^{-4}$ and the weight of the diffusive term $\alpha_{diff}$ is 5$e^{-10}$.The order of these weights is relatively lower because during our training process, we noticed that the convective term and diffusive term had orders of magnitude around $e^{4}$ and $e^{10}$, respectively, which were higher than the data loss component. In order to maintain a consistent order of magnitude across all components, we employed a trial-and-error approach to determine these weight values. This same rationale applies to our treatment of other datasets.
\begin{equation}
\begin{aligned}
&\mathcal{L}_{conv} = |(\mathbf{u} \cdot \nabla \mathbf{u})_{prediction}-(\mathbf{u} \cdot \nabla \mathbf{u})_{fine}|\\
&\mathcal{L}_{diff} = |(\nabla^2 \mathbf{u})_{pred}-(\nabla^2 \mathbf{u})_{fine}|
\end{aligned}
\end{equation}
\textbf{2D-Methane Combustion} we have set the weight for the convective term $\alpha_{conv}$ is 1$e^{-5}$ and the weight of the diffusive term $\alpha_{diff}$ is 1$e^{-9}$.The value of weight $\alpha_{conv}$ and $\alpha_{diff}$ are obtained by trial and error method.
\begin{equation}
\begin{aligned}
& \text{convective term} = \mathbf{u} \cdot \nabla \mathbf{u}+ \nabla \cdot (\mathbf{u} h + \mathbf{u}\frac{u^{2}}{2}) + \nabla \cdot \mathbf{u}Y_{i}\\
&\mathcal{L}_{conv} = |(\text{convective term})_{pred}-(\text{convective term})_{fine}|\\
&\mathcal{L}_{diff} = |(\nabla^2 \mathbf{u} + \nabla^2 h + \nabla^2 Y_i)_{pred}-(\nabla^2 \mathbf{u} + \nabla^2 h + \nabla^2 Y_i)_{fine}|
\end{aligned}
\end{equation}
\textbf{Industrial Heat Exchanger} we have set the weight for the convective term $\alpha_{conv}$ is 1$e^{-6}$ and the weight of the conductive term $\alpha_{cond}$ is 1$e^{-10}$.The value of weight $\alpha_{conv}$ and $\alpha_{cond}$ are obtained by trial and error method. 
\begin{equation}
    \hspace{-2.2cm}\begin{aligned}
     &\text{Physics Loss:}\hspace{1.8cm} \mathcal{L}_{physics} =  \alpha_{conv}(\mathcal{L}_{conv,m}+\mathcal{L}_{conv,f}) + \alpha_{cond}\mathcal{L}_{cond} 
    \end{aligned}
\end{equation}
\begin{equation}
\begin{aligned}
&\mathcal{L}_{conv,m} = \left|(\partial_{\varphi} T_m)_{prediction}-(\partial_{\varphi} T_m)_{fine}\right|\\
&\mathcal{L}_{conv,f} = \left|( \partial_z T)_{prediction}-(\partial_z T)_{fine}\right|\\
&\mathcal{L}_{cond} =  \left|(\partial_{z}^2 T_m )_{prediction}-(\partial_{z}^2 T_m)_{fine}\right|
\end{aligned}
\end{equation}

The derivatives of the governing equations in the physics loss were computed using a 2nd-order finite difference method (FDM). Although there is an intention to explore higher-order FDM techniques in the future for potentially improved resolution in data feature prediction.

This loss function formulation allows for the simultaneous optimization of data fidelity and adherence to physics constraints during the training of the model. 
\subsection{Comparative Analysis: Downsampling vs. Coarse Mesh Representation}
\label{sec:Downsampling}
\begin{figure}[htbp]
    \centering
    \includegraphics[scale=0.55]{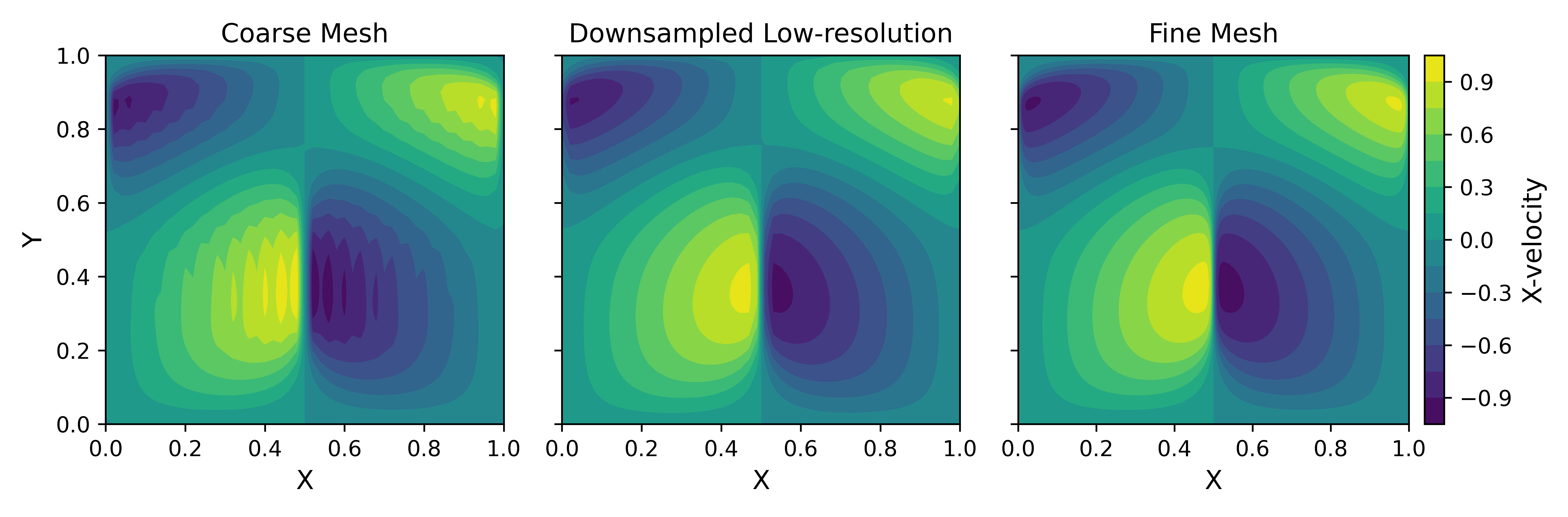}
    \caption{Comparison of Contour Surface Plots for $U_x$ in 2D-Burgers eq: Coarse Mesh vs. Downsampling vs. Fine Mesh Data}
    \label{fig:downsampling}
\end{figure}
\begin{figure}
    \centering
    \includegraphics[scale=0.55]{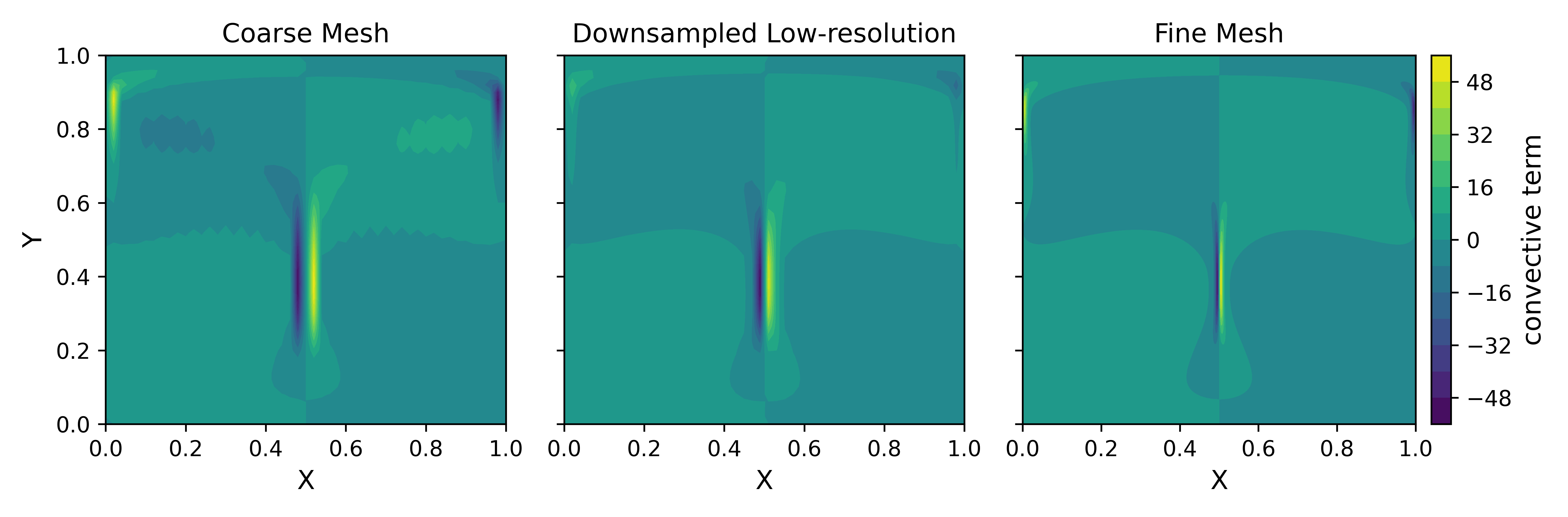}
    \caption{Comparison of Contour Surface Plots for $\mathbf{u} \cdot \nabla \mathbf{u}$ in 2D-Burgers eq: Coarse Mesh vs. Downsampling vs. Fine Mesh Data}
    \label{fig:convective term}
\end{figure}

\begin{table}[htbp]
\centering
\caption{Comparison of vanilla UNet performance on downsampled low-resolution data and coarse mesh data.}
\label{tab:downsampling}
\resizebox{0.8\textwidth}{!}{
\begin{tabular}{c|c|c|c|c}
\toprule
\textbf{Algorithm}    & \textbf{Data}                             & \textbf{RMSE}                & \textbf{MAE}                 & \textbf{R$^2$}                \\ \midrule
\multirow{2}{*}{UNet} & Downsampled low-resolution (51 $\times$ 51) & 0.02125                      & 0.00426                      & 0.9944                     \\ \cmidrule(l){2-5} 
                      & Coarse Mesh (51 $\times$ 51)                & 0.0283                       & 0.01265                      & 0.9955                     \\ \midrule
PIUNet                & Coarse Mesh (51 $\times$ 51)                & \multicolumn{1}{l|}{0.02075} & \multicolumn{1}{l|}{0.00628} & \multicolumn{1}{l}{0.9948} \\ \bottomrule
\end{tabular}
}
\end{table}
Figure \ref{fig:downsampling} demonstrates that generating low-resolution data by simply downsampling the high-resolution data retains the majority of the underlying physics. In this context, we applied average pooling to downsample the fine-mesh data (high-resolution) by a factor of 8, resulting in a low-resolution grid of 51x51, which matches the resolution of our corresponding coarse-grid data. Additionally, figure \ref{fig:convective term}, which examines the convective term ($\mathbf{u} \cdot \nabla \mathbf{u}$), further substantiates our claim. To facilitate a fair comparison, we also applied the vanilla UNet super-resolution technique to both the downsampled low-resolution data and the coarse mesh data. From the results presented in Table 1, it is evident that vanilla UNet is more successful at super-resolving the downsampled low-resolution data to the fine mesh level compared to the coarse mesh data. This outcome is attributed to the fact that the downsampled low-resolution data retains a higher degree of the governing physics from the high-resolution data. Remarkably, the UNet super-resolution result of the downsampled low-resolution data is on par with that of PIUNet.

\subsection{Additional plots}

\textbf{2D-Burger Equation}

\begin{figure}[htbp]
    \centering
    
    \begin{subfigure}{0.4\textwidth}
        \centering
        \includegraphics[scale=0.65]{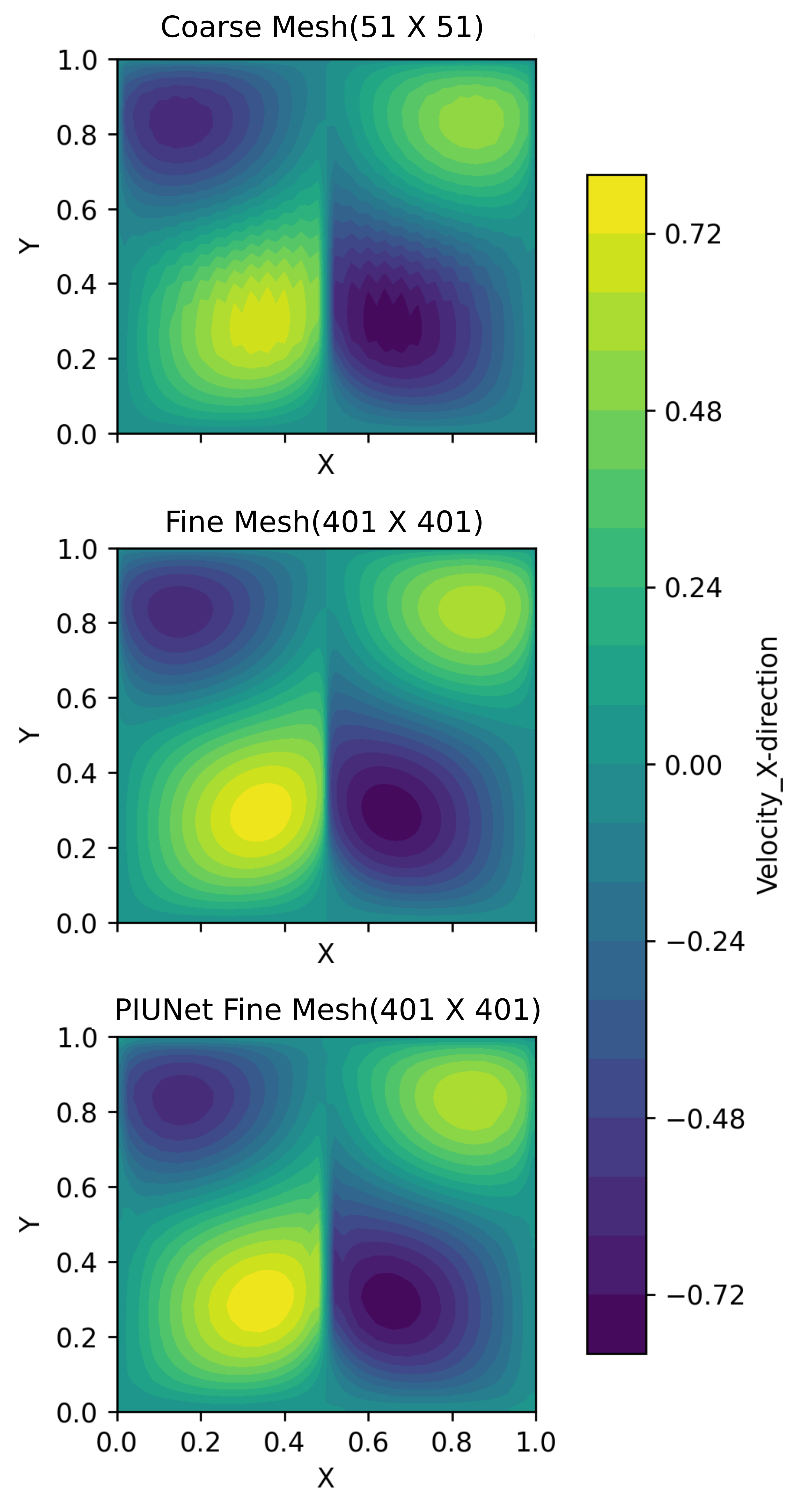}
        \caption{Contour Surface Plot of x-velocity}
    \end{subfigure}
    \hspace{1cm}
    \begin{subfigure}{0.4\textwidth}
        \centering
        \includegraphics[scale=0.47]{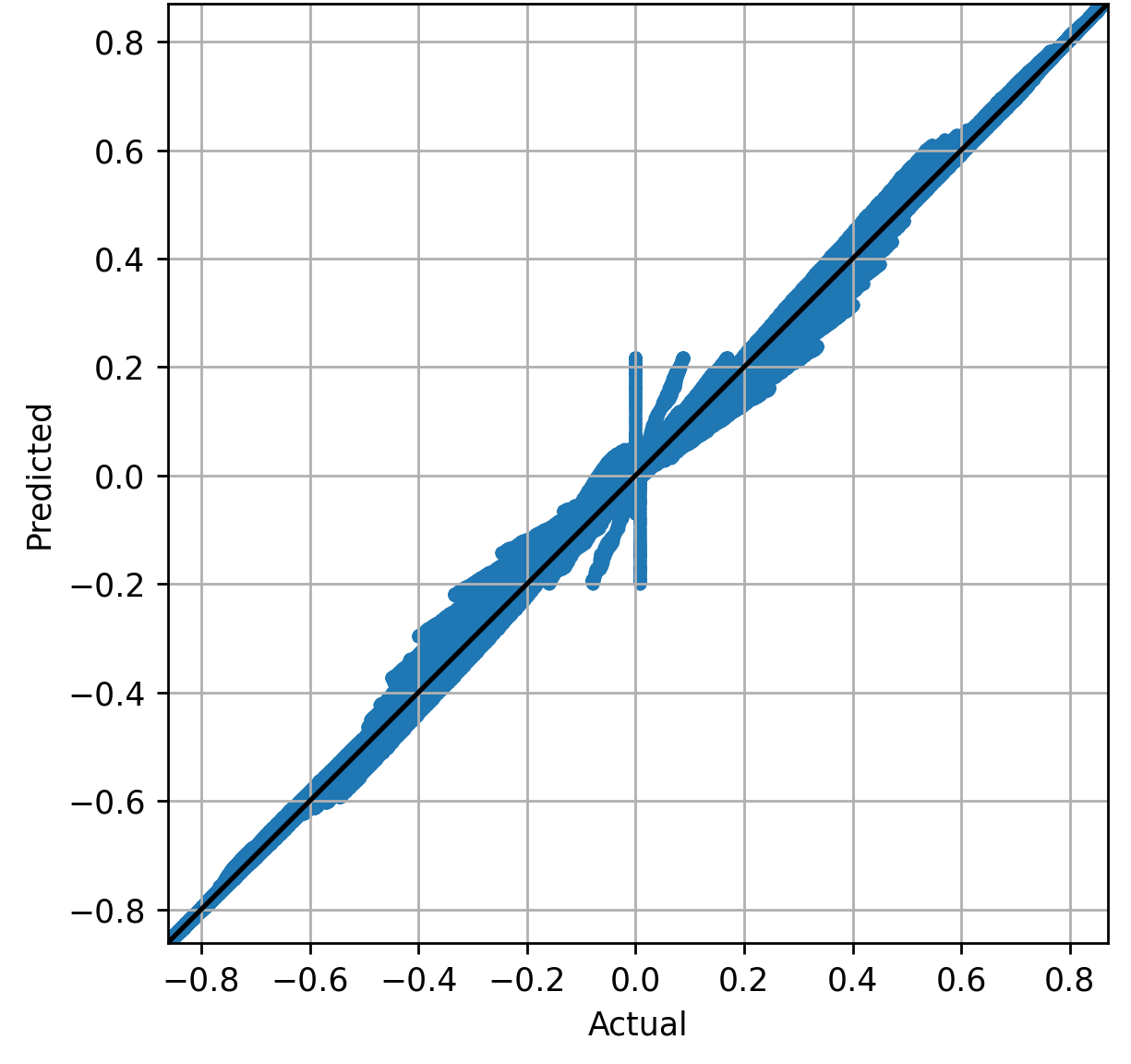}
        \caption{Parity plot}
        \includegraphics[scale=0.47]{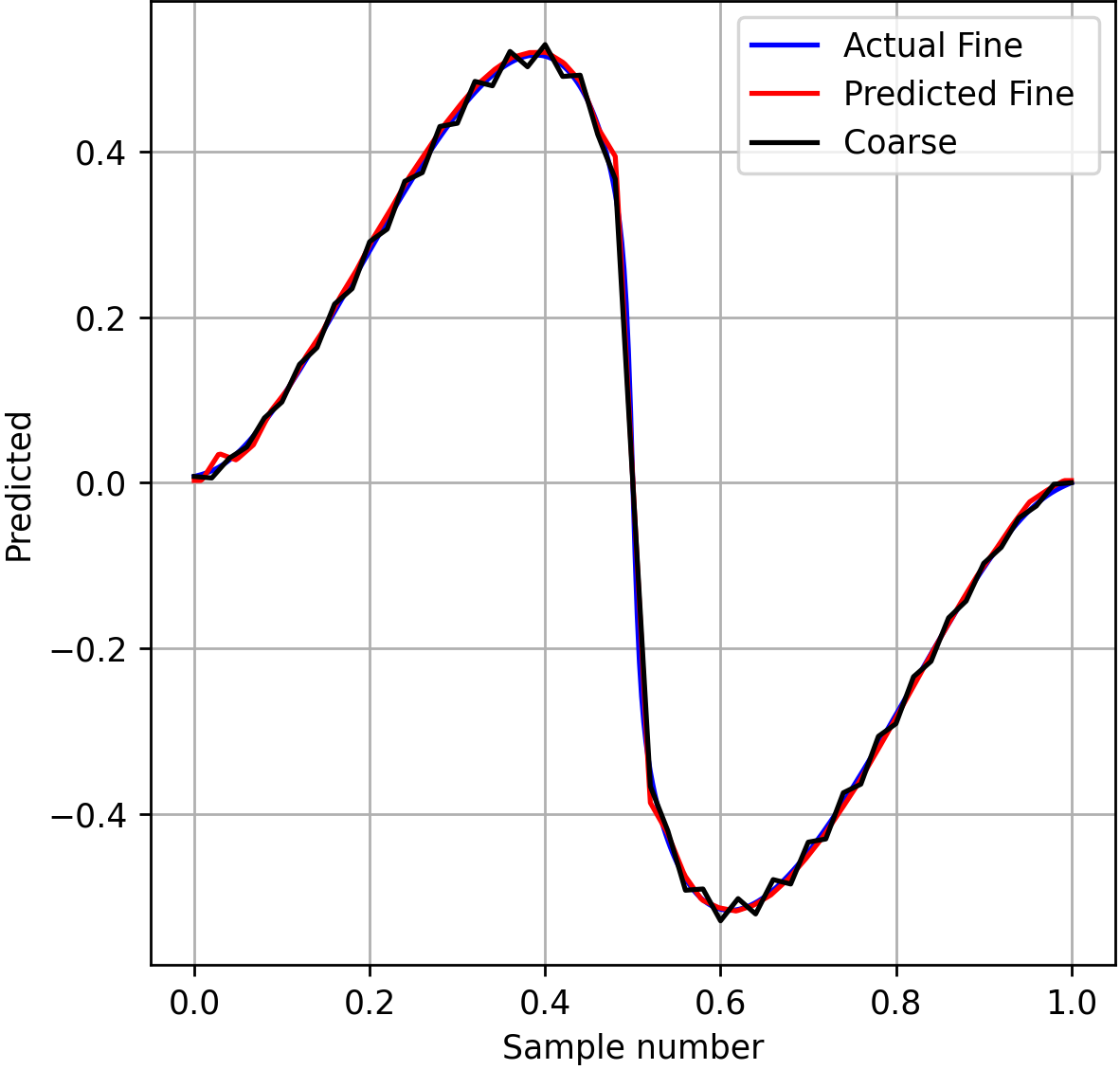}
        \caption{Centerline x-velocity trend plot}
    \end{subfigure}

    \caption{Velocity plots of 2D-Burger Equation }
    \label{fig:burger}
\end{figure}

In Figure \ref{fig:burger}, the contour surface plot provides a visual representation of PIUNet's proficiency in predicting the solution for the fine mesh grid (401×401) with a mean L2 error of 0.0097 when compared to the actual fine mesh data. Conversely, for the coarse mesh grid (51×51), the mean L2 error is 0.0446 in comparison to the actual fine mesh data, showcasing a significant reduction of approximately 78\% in the mean L2 error achieved through PIUNet's predictions. This reduction underscores the effectiveness of PIUNet in enhancing accuracy. Furthermore, upon examining the centerline trend plot of $U_x$, it becomes evident that PIUNet adeptly replicates the observed behavior present in the actual fine mesh data. However, it's worth noting that, particularly in the vicinity of the discontinuity region, there are some values with small mismatches between the actual and predicted data, as observed in the parity plot.

\newpage
\textbf{Methane Combustion}

\begin{figure}[htbp]
    \centering
    
    \begin{subfigure}{\textwidth}
        \centering
        \includegraphics[scale=0.7]{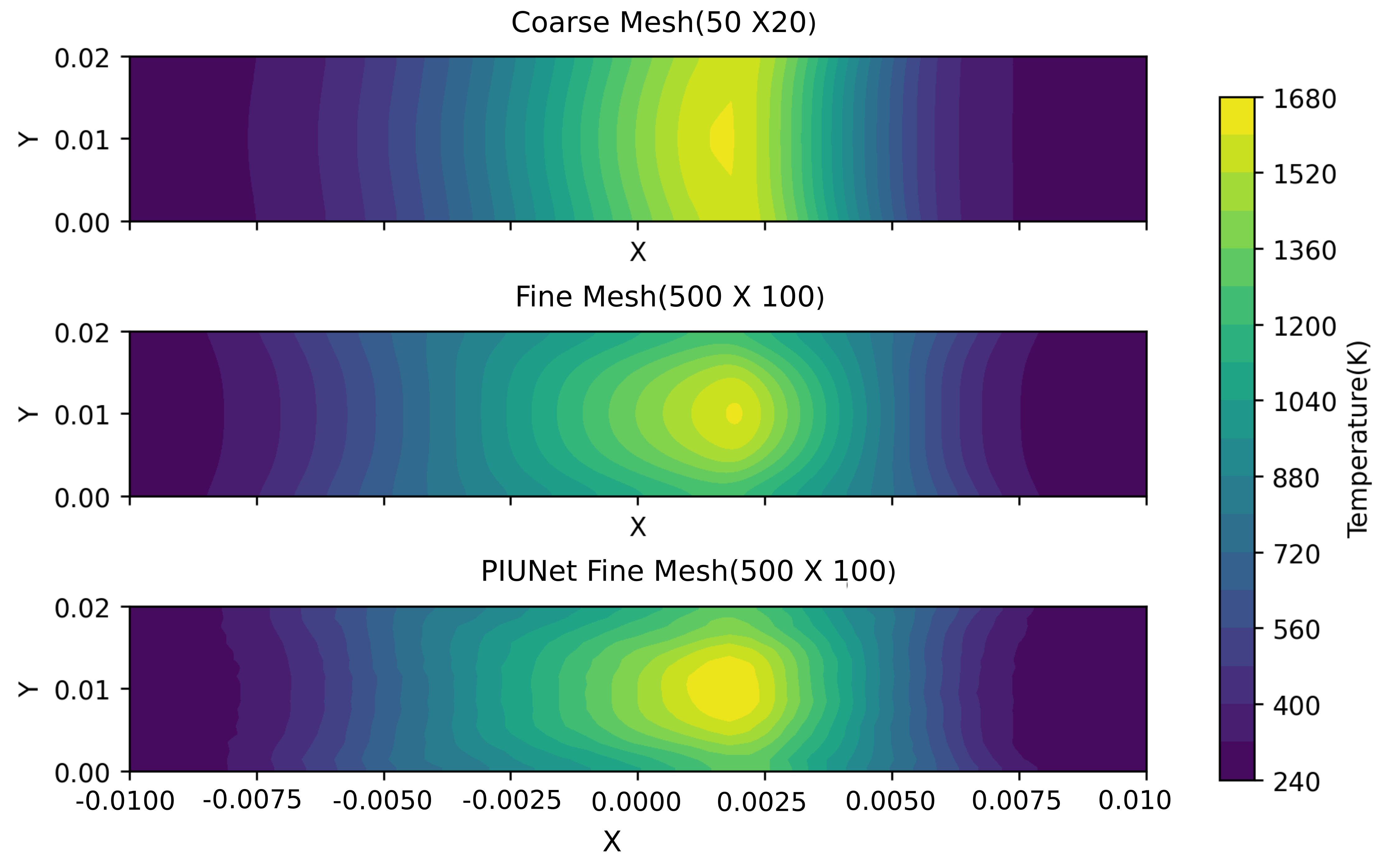}
        \caption{Contour surface plot of Adiabatic Temperature}
    \end{subfigure}

    \begin{subfigure}{0.45\textwidth}
        \centering
        \includegraphics[scale=0.42]{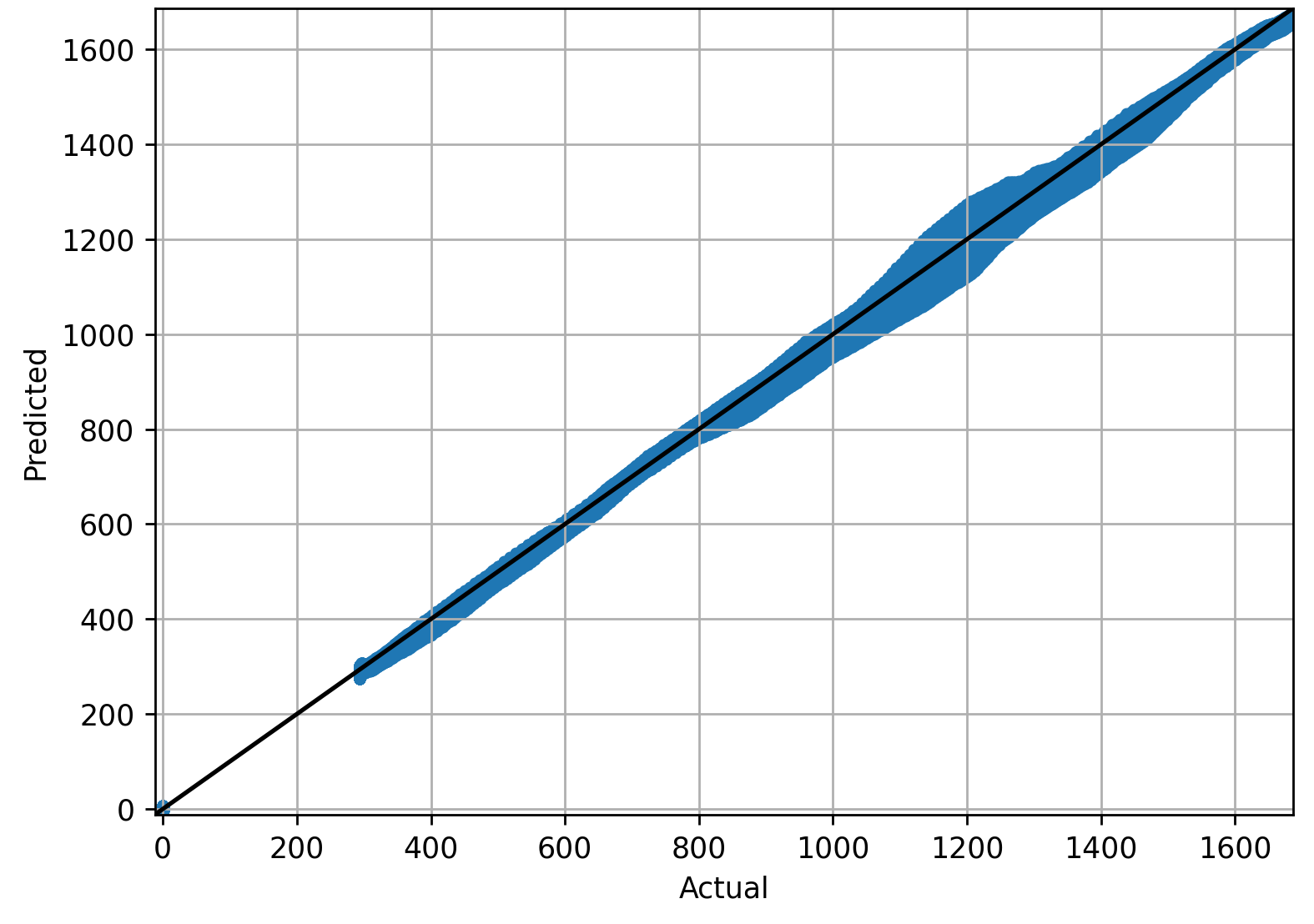}
        \caption{Parity plot}
    \end{subfigure}
    \hfill
    \begin{subfigure}{0.45\textwidth}
        \centering
        \includegraphics[scale=0.42]{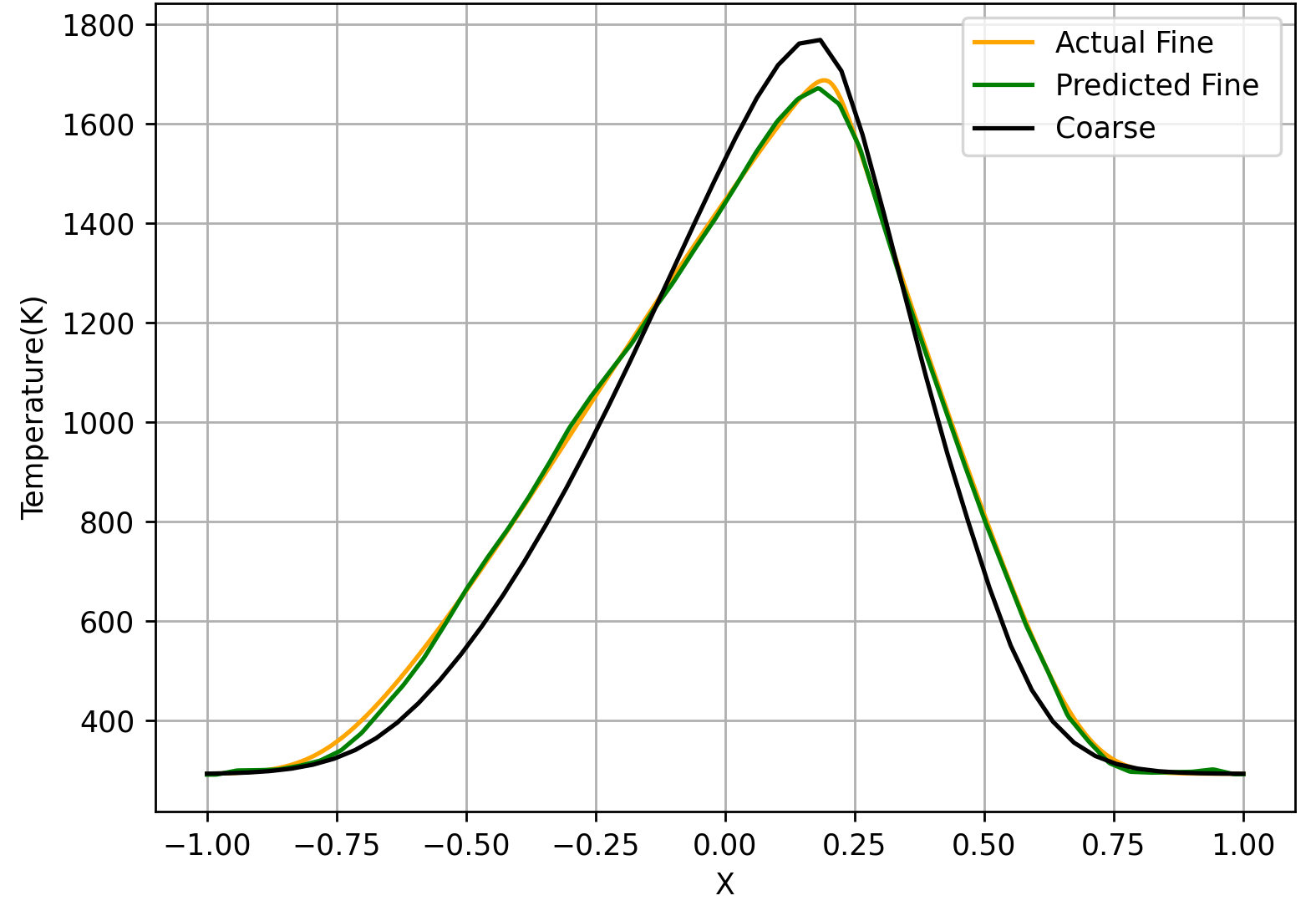}
        \caption{Centerline Temperature trend plot}
    \end{subfigure}
    
    \caption{Temperature plots of Methane combustion}
    \label{fig:2D_Methane}
\end{figure}

In the case of the methane combustion experiment, the adiabatic flame temperature holds significant importance. Contour surface plot, depicted in Figure \ref{fig:2D_Methane}, clearly demonstrates that PIUNet excels in capturing the gradient of each counter in the fine mesh simulation from coarse mesh data. Furthermore, the parity plot and the centerline temperature profile reveal that PIUNet accurately predicts the entire adiabatic flame temperature profile, including the peak temperature value.
\newpage
\textbf{Industrial Heat Exchanger}
\begin{figure}[htbp]
    \centering
    \begin{subfigure}{\textwidth}
        \centering
        \includegraphics[scale=0.6]{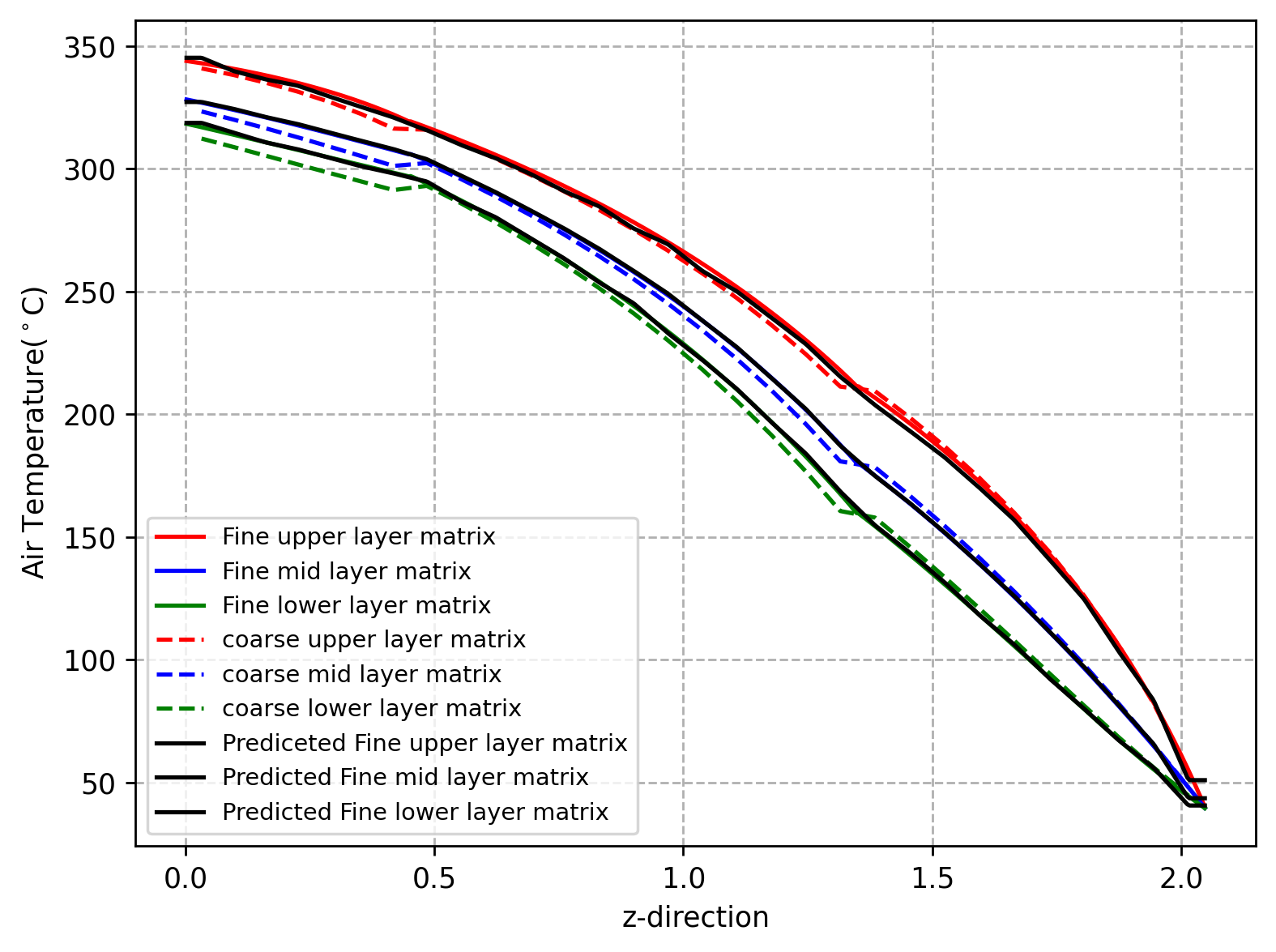}
        \caption{Air flow Temperature profile Plot}
    \end{subfigure}
    
    \begin{subfigure}{\textwidth}
        \centering
        \includegraphics[scale=0.6]{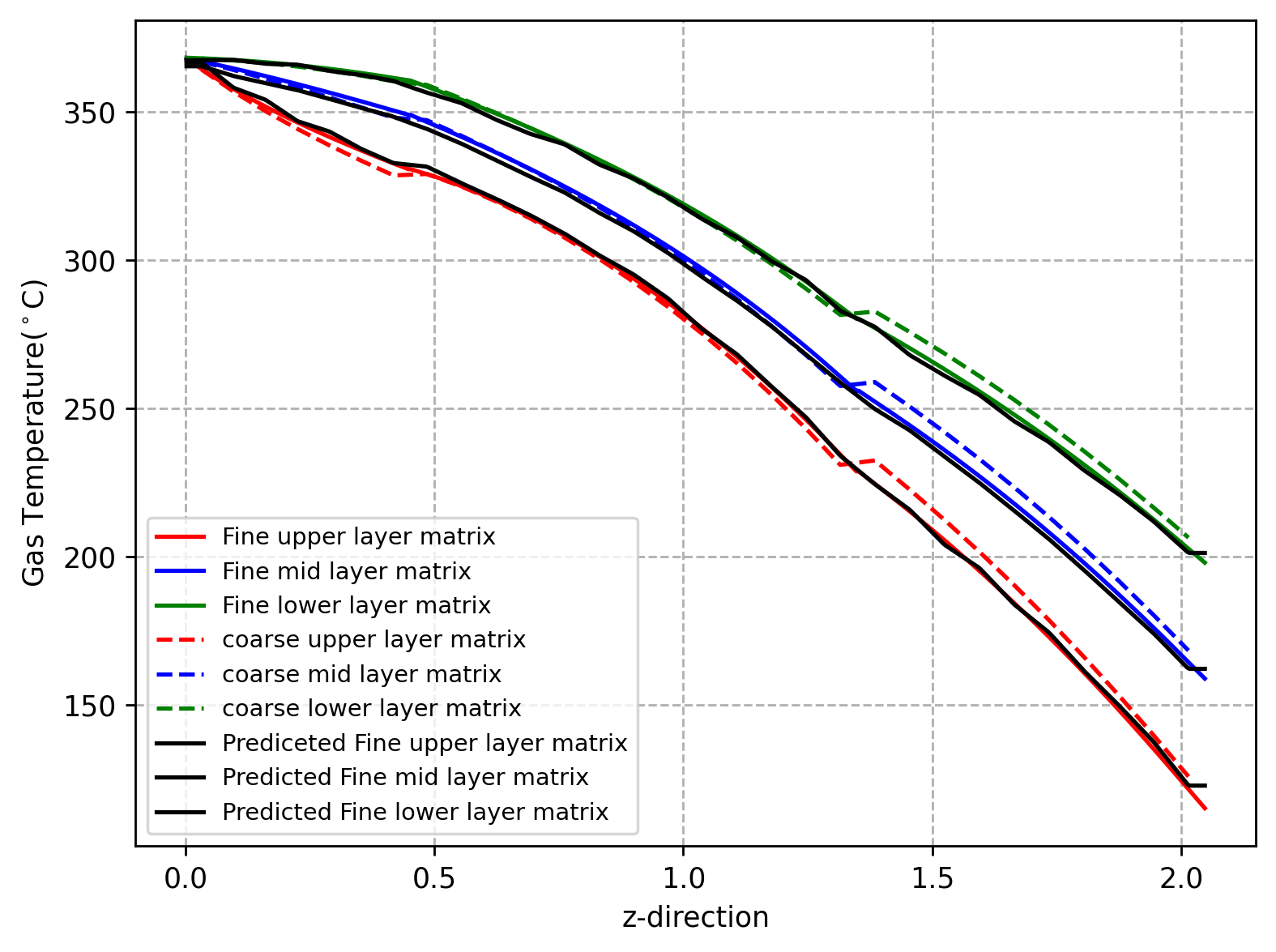}
        \caption{Gas flow Temperature profile Plot}
    \end{subfigure}
    
    \caption{Temperature profile comparison of coarse, fine and predicted fine mesh from PIUNet for Industrial Heat Exchanger}
\end{figure}

\end{document}